\documentclass[]{emulateapj}
\usepackage{epsfig, natbib, graphicx, color}
\usepackage[section]{placeins}
\usepackage{amsmath}
\usepackage{appendix}
\usepackage{multirow}

\input epsf 

\def\chisqredshift{103/30}
\def\chisqcolor{60/30}
\def\chisqstretch{42/30}
\def\chisqmb{55/30}

\def\ssigzspec{$0.000$}
\def\ssigzphot{$0.009$}

\def\shubblespec{$0.193$ mag}
\def\shubblephot{$0.201$ mag}

\def\dsigzspec{$0.000$}
\def\dsigzphot{$0.014$}

\def\dhubblespec{$0.196$ mag}
\def\dhubblephot{$0.193$ mag}

\def\sspecsigint{0.106}
\def\sphotsigint{0.111}

\def\dspecsigint{0.138}
\def\dphotsigint{0.145}

\def\sdeltawone{-0.0011}

\def\sdeltawthree{-0.0050}

\def\sdeltawerrone{0.0020}

\def\sdeltawerrthree{0.0024}

\def\sdeltawstdone{$0.0249$}

\def\sdeltawstdthree{$0.0294$}

\def\ddeltawone{0.0049}

\def\ddeltawthree{-0.0146}

\def\dwuncertzero{$0.0432$}
\def\dwuncertone{$0.0458$}

\def\dwuncertthree{$0.0641$}

\def\swuncertzero{$0.0753$}
\usepackage{color}
\usepackage{lineno}
\usepackage{hyperref}
\usepackage{array}
\usepackage{ulem}

\newcommand{\sigmadeltaz}{\sigma_{\Delta z/(1+z)}}

\usepackage[hang]{footmisc}

\shortauthors{Chen et al.}

\begin{document}

\begin{nolinenumbers}
\vspace*{-\headsep}\vspace*{\headheight}
\footnotesize \hfill FERMILAB-PUB-22-082-PPD\\
\vspace*{-\headsep}\vspace*{\headheight}
\footnotesize \hfill DES-2021-0665
\end{nolinenumbers}

\title{Measuring Cosmological Parameters with Type Ia Supernovae in redMaGiC Galaxies}

\def\andname{}

\author{
R.~Chen\altaffilmark{1},
D.~Scolnic\altaffilmark{1},
E.~Rozo\altaffilmark{2},
E.~S.~Rykoff\altaffilmark{3,4},
B.~Popovic\altaffilmark{1},
R.~Kessler\altaffilmark{5,6},
M.~Vincenzi\altaffilmark{1},
T.~M.~Davis\altaffilmark{7},
P.~Armstrong\altaffilmark{8},
D.~Brout\altaffilmark{9,10},
L.~Galbany\altaffilmark{11,12},
L.~Kelsey\altaffilmark{13,14},
C.~Lidman\altaffilmark{15,8},
A.~M\"oller\altaffilmark{16},
B.~Rose\altaffilmark{1},
M.~Sako\altaffilmark{17},
M.~Sullivan\altaffilmark{14},
G.~Taylor\altaffilmark{8},
P.~Wiseman\altaffilmark{14},
J.~Asorey\altaffilmark{18},
A.~Carr\altaffilmark{7},
C.~Conselice\altaffilmark{19,20},
K.~Kuehn\altaffilmark{21,22},
G.~F.~Lewis\altaffilmark{23},
E.~Macaulay\altaffilmark{13},
M.~Rodriguez-Monroy\altaffilmark{24},
B.~E.~Tucker\altaffilmark{8},
T.~M.~C.~Abbott\altaffilmark{25},
M.~Aguena\altaffilmark{26},
S.~Allam\altaffilmark{27},
F.~Andrade-Oliveira\altaffilmark{28},
J.~Annis\altaffilmark{27},
D.~Bacon\altaffilmark{13},
E.~Bertin\altaffilmark{29,30},
S.~Bocquet\altaffilmark{31},
D.~Brooks\altaffilmark{32},
D.~L.~Burke\altaffilmark{3,4},
A.~Carnero~Rosell\altaffilmark{33,26,34},
M.~Carrasco~Kind\altaffilmark{35,36},
J.~Carretero\altaffilmark{37},
R.~Cawthon\altaffilmark{38},
M.~Costanzi\altaffilmark{39,40,41},
L.~N.~da Costa\altaffilmark{26,42},
M.~E.~S.~Pereira\altaffilmark{43},
S.~Desai\altaffilmark{44},
H.~T.~Diehl\altaffilmark{27},
P.~Doel\altaffilmark{32},
S.~Everett\altaffilmark{45},
I.~Ferrero\altaffilmark{46},
B.~Flaugher\altaffilmark{27},
D.~Friedel\altaffilmark{35},
J.~Frieman\altaffilmark{27,6},
J.~Garc\'ia-Bellido\altaffilmark{47},
M.~Gatti\altaffilmark{17},
E.~Gaztanaga\altaffilmark{11,12},
D.~Gruen\altaffilmark{48,31},
S.~R.~Hinton\altaffilmark{7},
D.~L.~Hollowood\altaffilmark{45},
K.~Honscheid\altaffilmark{49,50},
D.~J.~James\altaffilmark{10},
O.~Lahav\altaffilmark{32},
M.~Lima\altaffilmark{51,26},
M.~March\altaffilmark{17},
F.~Menanteau\altaffilmark{35,36},
R.~Miquel\altaffilmark{52,37},
R.~Morgan\altaffilmark{53},
A.~Palmese\altaffilmark{54},
F.~Paz-Chinch\'{o}n\altaffilmark{35,55},
A.~Pieres\altaffilmark{26,42},
A.~A.~Plazas~Malag\'on\altaffilmark{56},
J.~Prat\altaffilmark{5,6},
A.~K.~Romer\altaffilmark{57},
A.~Roodman\altaffilmark{3,4},
E.~Sanchez\altaffilmark{18},
M.~Schubnell\altaffilmark{28},
S.~Serrano\altaffilmark{11,12},
I.~Sevilla-Noarbe\altaffilmark{18},
M.~Smith\altaffilmark{14},
M.~Soares-Santos\altaffilmark{28},
E.~Suchyta\altaffilmark{58},
G.~Tarle\altaffilmark{28},
D.~Thomas\altaffilmark{13},
C.~To\altaffilmark{49},
D.~L.~Tucker\altaffilmark{27},
and T.~N.~Varga\altaffilmark{59,60}
\\ \vspace{0.2cm} (DES Collaboration) \\
}

\affil{$^{1}$ Department of Physics, Duke University Durham, NC 27708, USA}
\affil{$^{2}$ Department of Physics, University of Arizona, Tucson, AZ 85721, USA}
\affil{$^{3}$ Kavli Institute for Particle Astrophysics \& Cosmology, P. O. Box 2450, Stanford University, Stanford, CA 94305, USA}
\affil{$^{4}$ SLAC National Accelerator Laboratory, Menlo Park, CA 94025, USA}
\affil{$^{5}$ Department of Astronomy and Astrophysics, University of Chicago, Chicago, IL 60637, USA}
\affil{$^{6}$ Kavli Institute for Cosmological Physics, University of Chicago, Chicago, IL 60637, USA}
\affil{$^{7}$ School of Mathematics and Physics, University of Queensland,  Brisbane, QLD 4072, Australia}
\affil{$^{8}$ The Research School of Astronomy and Astrophysics, Australian National University, ACT 2601, Australia}
\affil{$^{9}$ NASA Einstein Fellow}
\affil{$^{10}$ Center for Astrophysics $\vert$ Harvard \& Smithsonian, 60 Garden Street, Cambridge, MA 02138, USA}
\affil{$^{11}$ Institut d'Estudis Espacials de Catalunya (IEEC), 08034 Barcelona, Spain}
\affil{$^{12}$ Institute of Space Sciences (ICE, CSIC),  Campus UAB, Carrer de Can Magrans, s/n,  08193 Barcelona, Spain}
\affil{$^{13}$ Institute of Cosmology and Gravitation, University of Portsmouth, Portsmouth, PO1 3FX, UK}
\affil{$^{14}$ School of Physics and Astronomy, University of Southampton,  Southampton, SO17 1BJ, UK}
\affil{$^{15}$ Centre for Gravitational Astrophysics, College of Science, The Australian National University, ACT 2601, Australia}
\affil{$^{16}$ Centre for Astrophysics \& Supercomputing, Swinburne University of Technology, Victoria 3122, Australia}
\affil{$^{17}$ Department of Physics and Astronomy, University of Pennsylvania, Philadelphia, PA 19104, USA}
\affil{$^{18}$ Centro de Investigaciones Energ\'eticas, Medioambientales y Tecnol\'ogicas (CIEMAT), Madrid, Spain}
\affil{$^{19}$ Jodrell Bank Center for Astrophysics, School of Physics and Astronomy, University of Manchester, Oxford Road, Manchester, M13 9PL, UK}
\affil{$^{20}$ University of Nottingham, School of Physics and Astronomy, Nottingham NG7 2RD, UK}
\affil{$^{21}$ Australian Astronomical Optics, Macquarie University, North Ryde, NSW 2113, Australia}
\affil{$^{22}$ Lowell Observatory, 1400 Mars Hill Rd, Flagstaff, AZ 86001, USA}
\affil{$^{23}$ Sydney Institute for Astronomy, School of Physics, A28, The University of Sydney, NSW 2006, Australia}
\affil{$^{24}$ Laboratoire de physique des 2 infinis Ir\`ene Joliot-Curie, CNRS Universit\'e Paris-Saclay, B\^at. 100, Facult\'e des sciences, F-91405 Orsay Cedex, France}
\affil{$^{25}$ Cerro Tololo Inter-American Observatory, NSF's National Optical-Infrared Astronomy Research Laboratory, Casilla 603, La Serena, Chile}
\affil{$^{26}$ Laborat\'orio Interinstitucional de e-Astronomia - LIneA, Rua Gal. Jos\'e Cristino 77, Rio de Janeiro, RJ - 20921-400, Brazil}
\affil{$^{27}$ Fermi National Accelerator Laboratory, P. O. Box 500, Batavia, IL 60510, USA}
\affil{$^{28}$ Department of Physics, University of Michigan, Ann Arbor, MI 48109, USA}
\affil{$^{29}$ CNRS, UMR 7095, Institut d'Astrophysique de Paris, F-75014, Paris, France}
\affil{$^{30}$ Sorbonne Universit\'es, UPMC Univ Paris 06, UMR 7095, Institut d'Astrophysique de Paris, F-75014, Paris, France}
\affil{$^{31}$ Faculty of Physics, Ludwig-Maximilians-Universit\"at, Scheinerstr. 1, 81679 Munich, Germany}
\affil{$^{32}$ Department of Physics \& Astronomy, University College London, Gower Street, London, WC1E 6BT, UK}
\affil{$^{33}$ Instituto de Astrofisica de Canarias, E-38205 La Laguna, Tenerife, Spain}
\affil{$^{34}$ Universidad de La Laguna, Dpto. Astrofísica, E-38206 La Laguna, Tenerife, Spain}
\affil{$^{35}$ Center for Astrophysical Surveys, National Center for Supercomputing Applications, 1205 West Clark St., Urbana, IL 61801, USA}
\affil{$^{36}$ Department of Astronomy, University of Illinois at Urbana-Champaign, 1002 W. Green Street, Urbana, IL 61801, USA}
\affil{$^{37}$ Institut de F\'{\i}sica d'Altes Energies (IFAE), The Barcelona Institute of Science and Technology, Campus UAB, 08193 Bellaterra (Barcelona) Spain}
\affil{$^{38}$ Physics Department, William Jewell College, Liberty, MO, 64068}
\affil{$^{39}$ Astronomy Unit, Department of Physics, University of Trieste, via Tiepolo 11, I-34131 Trieste, Italy}
\affil{$^{40}$ INAF-Osservatorio Astronomico di Trieste, via G. B. Tiepolo 11, I-34143 Trieste, Italy}
\affil{$^{41}$ Institute for Fundamental Physics of the Universe, Via Beirut 2, 34014 Trieste, Italy}
\affil{$^{42}$ Observat\'orio Nacional, Rua Gal. Jos\'e Cristino 77, Rio de Janeiro, RJ - 20921-400, Brazil}
\affil{$^{43}$ Hamburger Sternwarte, Universit\"{a}t Hamburg, Gojenbergsweg 112, 21029 Hamburg, Germany}
\affil{$^{44}$ Department of Physics, IIT Hyderabad, Kandi, Telangana 502285, India}
\affil{$^{45}$ Santa Cruz Institute for Particle Physics, Santa Cruz, CA 95064, USA}
\affil{$^{46}$ Institute of Theoretical Astrophysics, University of Oslo. P.O. Box 1029 Blindern, NO-0315 Oslo, Norway}
\affil{$^{47}$ Instituto de Fisica Teorica UAM/CSIC, Universidad Autonoma de Madrid, 28049 Madrid, Spain}
\affil{$^{48}$ Excellence Cluster Origins, Boltzmannstr.\ 2, 85748 Garching, Germany}
\affil{$^{49}$ Center for Cosmology and Astro-Particle Physics, The Ohio State University, Columbus, OH 43210, USA}
\affil{$^{50}$ Department of Physics, The Ohio State University, Columbus, OH 43210, USA}
\affil{$^{51}$ Departamento de F\'isica Matem\'atica, Instituto de F\'isica, Universidade de S\~ao Paulo, CP 66318, S\~ao Paulo, SP, 05314-970, Brazil}
\affil{$^{52}$ Instituci\'o Catalana de Recerca i Estudis Avan\c{c}ats, E-08010 Barcelona, Spain}
\affil{$^{53}$ Physics Department, 2320 Chamberlin Hall, University of Wisconsin-Madison, 1150 University Avenue Madison, WI  53706-1390}
\affil{$^{54}$ Department of Astronomy, University of California, Berkeley,  501 Campbell Hall, Berkeley, CA 94720, USA}
\affil{$^{55}$ Institute of Astronomy, University of Cambridge, Madingley Road, Cambridge CB3 0HA, UK}
\affil{$^{56}$ Department of Astrophysical Sciences, Princeton University, Peyton Hall, Princeton, NJ 08544, USA}
\affil{$^{57}$ Department of Physics and Astronomy, Pevensey Building, University of Sussex, Brighton, BN1 9QH, UK}
\affil{$^{58}$ Computer Science and Mathematics Division, Oak Ridge National Laboratory, Oak Ridge, TN 37831}
\affil{$^{59}$ Max Planck Institute for Extraterrestrial Physics, Giessenbachstrasse, 85748 Garching, Germany}
\affil{$^{60}$ Universit\"ats-Sternwarte, Fakult\"at f\"ur Physik, Ludwig-Maximilians Universit\"at M\"unchen, Scheinerstr. 1, 81679 M\"unchen, Germany}

\begin{abstract}
Current and future cosmological analyses with Type Ia Supernovae (SNe Ia) face three critical challenges: i) measuring redshifts from the supernova or its host galaxy; ii) classifying SNe without spectra; and iii) accounting for correlations between the properties of SNe Ia and their host galaxies. We present here a novel approach that addresses each challenge. In the context of the Dark Energy Survey (DES), we analyze a SNIa sample with host galaxies in the redMaGiC galaxy catalog, a selection of Luminous Red Galaxies. Photo-$z$ estimates for these galaxies are expected to be accurate to $\sigmadeltaz \sim0.02$. The DES-5YR photometrically classified SNIa sample contains approximately 1600 SNe and 125 of these SNe are in redMaGiC galaxies. We demonstrate that redMaGiC galaxies almost exclusively host SNe Ia, reducing concerns with classification uncertainties. With this subsample, we find similar Hubble scatter (to within $\sim0.01$ mag) using photometric redshifts in place of spectroscopic redshifts. With detailed simulations, we show the bias due to using photo-$z$s from redMaGiC host galaxies on the measurement of the dark energy equation-of-state $w$ is up to $\Delta w \sim 0.01-0.02$.  With real data, we measure a difference in $w$ when using redMaGiC photometric redshifts versus spectroscopic redshifts of $\Delta w = 0.005$. Finally, we discuss how SNe in redMaGiC galaxies appear to be a more standardizable population due to a weaker relation between color and luminosity ($\beta$) compared to the DES-3YR population by $\sim5\sigma$; this finding is consistent with predictions that redMaGiC galaxies exhibit lower reddening ratios ($\textrm{R}_\textrm{V}$) than the general population of SN host galaxies. These results establish the feasibility of performing redMaGiC SN cosmology with photometric survey data in the absence of spectroscopic data.
\end{abstract}

\section{Introduction}
Type Ia Supernovae (SNe Ia) remain a critical tool as standardizable candles to measure cosmological parameters and constrain models for dark energy.  In the next decade, multiple surveys such as the Vera Rubin Observatory Legacy Survey of Space and Time (LSST; \citealp{Ivezic19}) and the Nancy Grace Roman Space Telescope (Roman; \citealp{Hounsell18, NGRST}) will discover more than a million SNe which will be leveraged to make more precise measurements of the dark energy equation-of-state parameter ($w$) and its dependence on cosmic time. The success of these programs requires i) information about the SN type and ii) accurate determination of redshifts. The primary cosmological results from SN surveys have historically been reliant on spectroscopic information, including the first results from the Dark Energy Survey (DES-3YR; \citealp{DES3YR}), which were obtained from a sample of 207 spectroscopically-confirmed SNe Ia with available host-galaxy or SN redshifts. For totals of SNe approaching 2.4 million \citep{LSSTSRD, Frohmaier21}, it is impossible to spectroscopically observe each SN due to cost and time constraints. For this analysis, we present a novel solution to this problem by focusing on the sample of SNe Ia in a subset of galaxies where SN type and redshift can more easily be determined than in the general population.

For photometric classification of the SNe, recent analyses have made significant progress in rejecting core-collapse SNe (SNe Ibc and II) and selecting samples $>90\%$ pure. The Photometric LSST Astronomical Time-series Classification Challenge (PLAsTiCC; \citealp{PLAsTiCC18}) included a mix of 18 transient models \citep{Kessler19}, and the top performing light-curve classifiers achieved $95\%$ levels of purity by training on a subset of the data \citep{Hlozek20}. SuperNNova (SNN; \citealp{Moller20}), a neural net classifier trained on simulations that use PLAsTiCC models (SNIa, SNIax, SNIa-91bg, SNII, SNIb/c), has a predicted efficiency from DES simulations of 97.7-99.5\% \citep{Vincenzi21b, Moller22}. An alternate approach to photometric classification is to use host-galaxy information to avoid problems with low Signal-to-Noise Ratio (SNR) data and sparse sampling. \citet{Foley13} found that galaxy morphology provides the most discriminating information for determining a SNIa classification probability. Core-collapse SNe have massive ($>8 M_{\odot}$) star progenitors, consistent with observations that they explode almost exclusively in gas rich, star forming galaxies, whereas SNe Ia have white dwarf progenitors and appear in a variety of host-galaxy types.

To precisely measure redshifts, large-area surveys such as DES \citep{DES-all} and Pan-STARRS1 (PS1; \citealp{Chambers16}) have pursued dedicated host-galaxy follow-up programs. PS1 used the MMT Observatory and AAOmega spectrograph on the Anglo-Australian Telescope (AAT) to measure spectroscopic redshifts after the survey was completed.  DES had a concurrent program (OzDES; \citealp{Lidman20}) to measure redshifts during the survey. OzDES also used the AAOmega spectrograph on AAT, as the Two Degree Field system (2dF) + AAOmega has a similar field-of-view to the Dark Energy Camera (DECam). However, this approach to obtaining redshifts requires large amounts of dedicated telescope time and additional modeling of spectroscopic efficiency due to biases toward brighter host galaxies \citep{Vincenzi21a}. 

So far, there have been limited studies on using photometric redshift estimates (photo-$z$) in a cosmological study with SNe Ia. \citet{Kessler10a} produced LSST simulations and showed that a light-curve fit using a host-galaxy photo-$z$ prior yields comparable redshift precision to spectroscopic redshifts.  However, \citet{Sako11} found that using SN-only photo-$z$s with real SDSS data resulted in pathologies that propagated to biases in the distances and therefore the measurement of cosmological parameters. Other studies \citep{Wojtak15, Davis19} have found that systematic redshift errors as small as $10^{-4}$ can mimic a 1\% perturbation in $w$ but also illustrate that the impact of redshift biases diminishes with increasing redshift.

For SNIa samples with diverse host-galaxy types, the issue of preferentially targeting brighter galaxies is particularly problematic because there is a correlation between the mass and rest-frame U-R color of the galaxies and the luminosity of the SNe \citep{Sullivan10, Kelly10, Kelsey20}. There have also been observed correlations between other global host-galaxy properties such as metallicity and morphological type \citep{Hamuy00, Kelly10, Smith20}, as well as local host-galaxy environments \citep{Rigault13, Rigault15, Rigault20, Roman18, Kelsey20}. These correlations are not well understood and are the subject of ongoing efforts to implement better bias corrections and modeling \citep{Smith20, Rigault20, BS20, Popovic21}.  Since the measurement of $w$ is based on a relative measurement between distances of SNe at high- and low-redshift, a redshift dependent selection of galaxy type may cause a significant systematic in measurements of $w$.

Here we investigate a solution to the above problems by exploiting SNe located in Luminous Red Galaxies (LRGs), which are a well known homogeneous population consisting of so called ``red and dead" elliptical galaxies. LRGs are expected to contain very low rates of core-collapse SNe, as core-collapse progenitors are massive and largely present in active star forming galaxies such as spiral galaxies. \citet{Foley13} found that 98\% of all SNe with elliptical host galaxies in the Lick Observatory Supernova Search sample \citep{Leaman11} are SNe Ia, implying that host-galaxy information alone can reduce photometric contamination from core-collapse SNe, and \citet{Irani21} conclude that only $0.3\%^{+0.3}_{-0.1}$ of all core-collapse SNe have elliptical hosts. Secondly, LRG spectra contain a prominent 4000 \r{A} break, which enables precise and accurate photo-$z$ estimates that have traditionally been utilized in large-scale structure studies. The photo-$z$ bias of these galaxies can be further constrained with the use of red galaxies selected using the redMaGiC algorithm described in \citet{Rozo16}, which utilizes a modified photo-$z$ estimator based on a full red-sequence model. Lastly, limiting the host-galaxy type allows for a more consistent sample across redshifts that is less sensitive to complicated correlations between SN light-curve properties and host-galaxy properties. While in the future there will be enough low-redshift SNe observed in LRGs to have a SN sample solely in one host-galaxy type, this study instead combines SNe in LRGs at $z > 0.05$ with the traditional spectroscopic low-redshift sample to provide a SN sample large enough for Ia cosmological analysis.

Less than 10\% of SN host galaxies are expected to be LRGs \citep{Foley13}. For DES, which has 3,627 SNe Ia before light-curve quality cuts and 1,606 after quality cuts, this LRG selection yields 227 SNe before cuts and 125 after cuts (6.26$\%$ and 7.78$\%$ respectively).  For the LSST sample of $\sim$2 million SNe, we expect $\sim 10^{5}$ SNe in LRGs.

In this paper, we provide a first investigation into the feasibility of using photometric redshifts from the redMaGiC galaxy catalog in a Type Ia Supernova cosmology analysis. The outline of the paper is as follows.  In Section \ref{sec:data}, we summarize the DES SN and host-galaxy data used for this work.  In Section \ref{sec:sims}, we discuss the simulations used to validate the method and for estimating bias corrections.  In Section \ref{sec:analysis}, we discuss the application of the method to DES-5YR data.  In Section \ref{sec:disc}, we discuss the implications and results of our study. In Section \ref{sec:conc}, we present our conclusions.

\section{Data}\label{sec:data}

\begin{figure*}[!hbt]
    \centering
	\includegraphics[scale=.58]{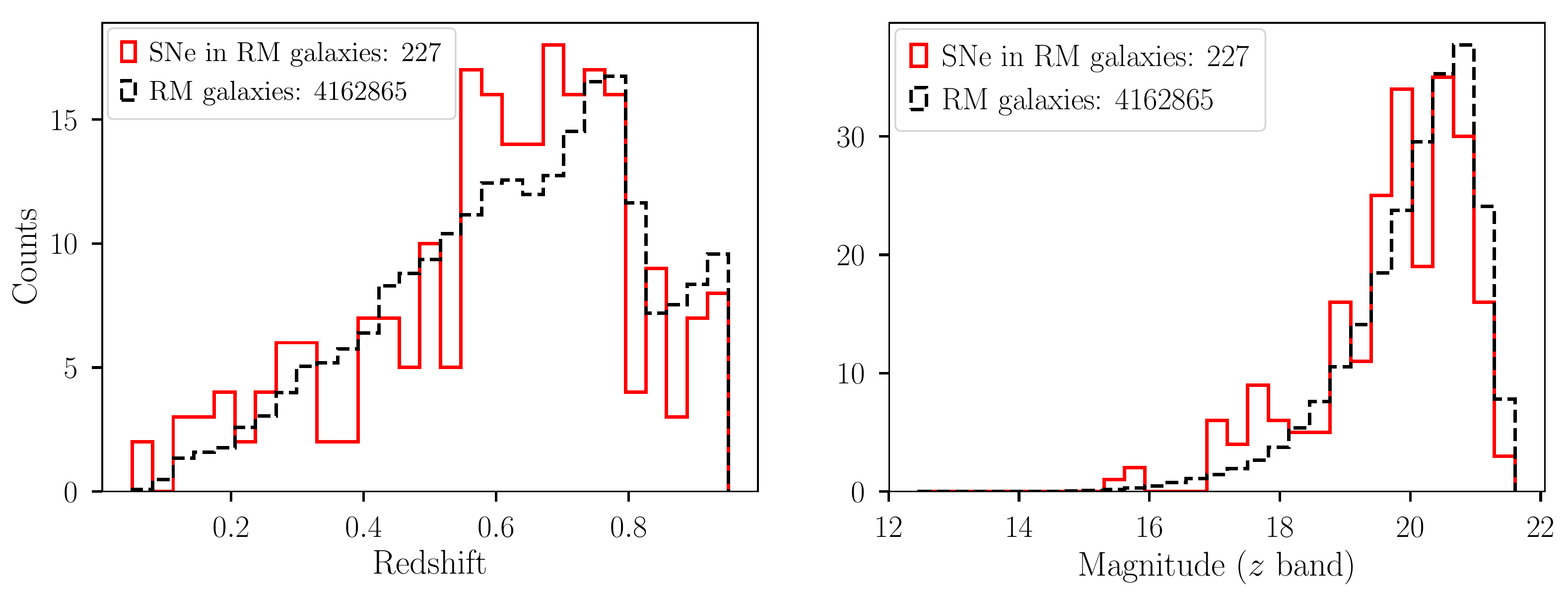}
	\caption{\textbf{Left:} The distribution of redMaGiC (RM) photometric redshifts for all redMaGiC galaxies in dashed black compared to the subset with a matched SN in solid red. \textbf{Right: } The distribution of $z$ band magnitudes for all redMaGiC galaxies compared to the subset with a matched SN. The histogram for all RM galaxies is scaled to match the sum of RM galaxies that host SNe.}
	\label{fig:redshift_mag_dists}
\end{figure*}

\subsection{redMaGiC Galaxies}

Luminous Red Galaxies occupy a very narrow range in color and intrinsic luminosity, and they contain old, red stellar populations. LRGs are a useful probe for large-scale structure studies \citep{Stoughton02}, as they are intrinsically luminous and therefore can be observed out to high redshift, and they are relatively massive and therefore tend to cluster strongly. LRG spectral energy distributions (SEDs) have a prominent 4000 \r{A} feature caused by absorption lines from metals in stellar atmospheres that make LRGs ideal candidates for photo-$z$ estimations. 

The red galaxies and photo-$z$ data used in this study were obtained using the red-sequence Matched-filter Galaxy (redMaGiC) algorithm \citep{Rozo16} run on six seasons of DES data with preliminary photometry. The algorithm is based on the infrastructure of redMaPPer \citep{Rykoff14}, a red sequence cluster finder designed for large photometric surveys. The algorithm selects red galaxies based on a chosen comoving space density and luminosity threshold. First, redMaGiC fits every red sequence galaxy with the redMaPPer red sequence template and computes its best fit photo-$z$. Using this photo-$z$, it computes the galaxy luminosity. Lastly, it applies selection requirements (cuts) on the luminosity and $\chi^2$ of the template fit, with the cuts tuned to select a desired comoving space density.

After applying redMaGiC cuts, a subset of galaxies with spectroscopic redshifts that are members of redMaPPer clusters is used to further train and validate a photo-$z$ afterburner. To avoid biased selection from using galaxies with spectroscopic follow-up, the redshift calibration uses redMaPPer photometric cluster redshifts ($z_{\textrm{cluster}}$) where each $z_{\textrm{cluster}}$ is fit simultaneously with all cluster members and is therefore more accurate than any individual galaxy redshift. The training sample's median redshift offset $z_{\textrm{cluster}} - z_{\textrm{red}}$ is calculated in bins of $z_{\textrm{red}}$, where $z_{\textrm{cluster}}$ is the cluster redshift and $z_{\textrm{red}}$ is the initial photometric redshift. This median offset is added to $z_{\textrm{red}}$ using spline interpolation to give the final photometric redshift, $z_{\textrm{redmagic}}$. 

The redshift range of the sample is $z_{\textrm{redmagic}} \in [0.05,0.95]$ and is shown in Figure \ref{fig:redshift_mag_dists}, along with the $z$ band magnitude ($m_z$) distribution. The redshift distribution peaks around $z = 0.7$ and the magnitude distribution peaks around $m_z = 21.0$. The redMaGiC algorithm is typically run to produce two sets of catalogs: ``high-luminosity" and ``high-density." The high-luminosity catalog restricts selection to galaxies with luminosity greater than 1 $L_{*}$ (as defined in \citealp{Rykoff16}) to extend to the highest redshift possible. The high-density catalog requires density of $\sim 10^{-3}$ $\textrm{Mpc}^{-3}$ with luminosity threshold 0.5 $L_{*}$. To use the largest possible selection of galaxies, we combine the two catalogs to create our catalog of redMaGiC galaxies. To compare the performance of spectroscopic redshifts ($z_{\textrm{spec}}$) and $z_{\textrm{redmagic}}$, we select galaxies with an available spectroscopic redshift, which reduces the number of galaxies from 4,162,865 to 55,735.

\begin{figure*}[htb!]
	\centering
	\includegraphics[scale=.49]{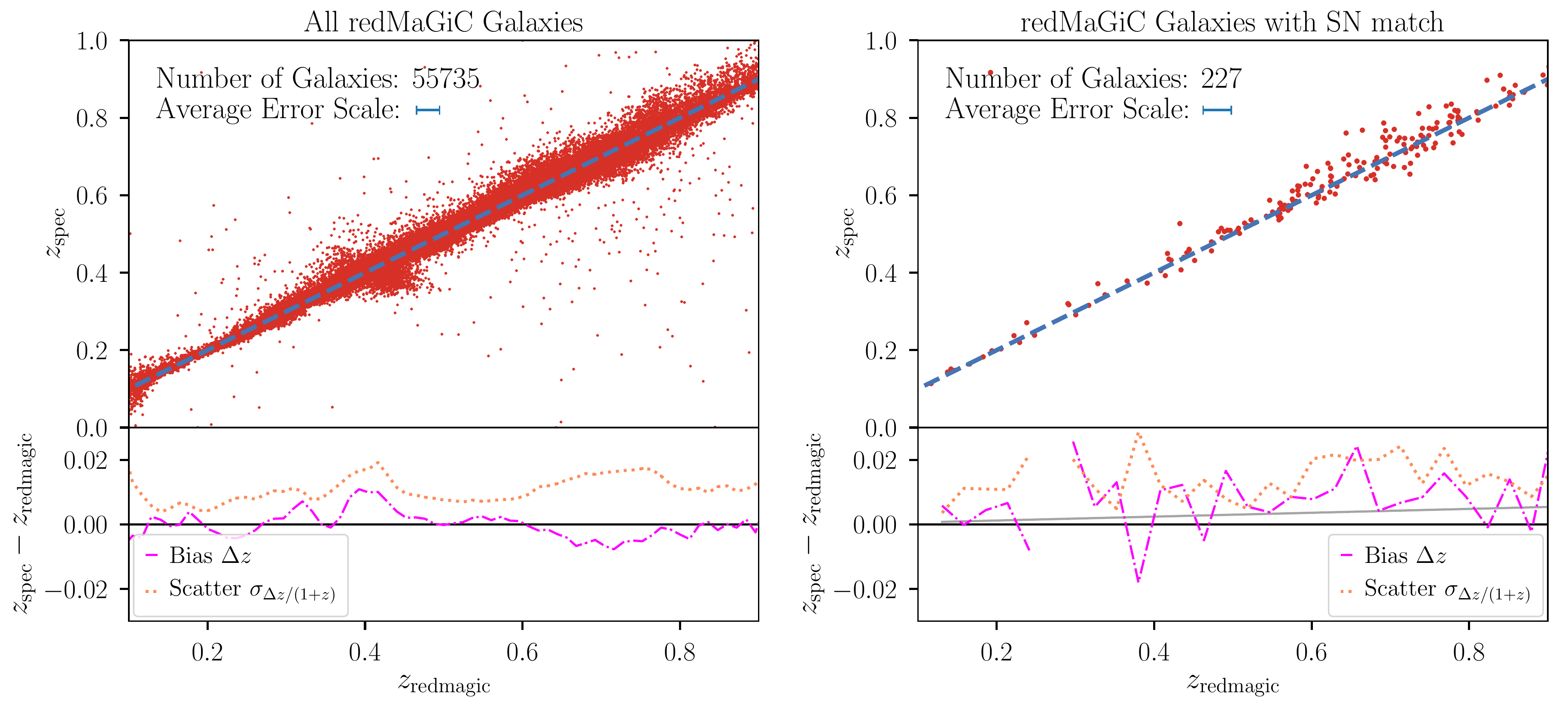}
	\caption{\textbf{Top Left:} Photometric redshifts vs.\ spectroscopic redshifts of redMaGiC galaxies. The dashed line has a slope of 1 for visual comparison. \textbf{Bottom Left:} Bias (binned $\overline{\Delta z}$) in dash-dotted pink and scatter in dotted yellow for redMaGiC photo-$z$. Only galaxies with a spectroscopic redshift are included in the bias and scatter calculation. \textbf{Right:} Same as left, but for the subset of redMaGiC galaxies with a SN match. A bias of $0.006z$ is shown in grey in the bottom panel.}
	\label{fig:RM_zbias}
\end{figure*}

To quantify the performance of photometric redshifts, the photo-$z$ bias $\overline{\Delta z}$ is defined as the median of offsets $\Delta z = z_{\textrm{spec}} - z_{\textrm{redmagic}}$, and the photo-$z$ scatter $\sigma_{\Delta z/(1+z)}$ is defined as $1.4826 \times \textrm{MAD}$, where MAD is the median absolute deviation $\mid \Delta z - \overline{\Delta z} \mid / (1+z_{\textrm{spec}})$. Figure \ref{fig:RM_zbias} shows the photo-$z$ performance of the combined high-density and high-luminosity Year 6 redMaGiC sample. A comparison to a one-to-one relation between spec-$z$ and photo-$z$ can be seen in the upper halves of the plots. The bottom halves of the plots show the bias and scatter plotted as a function of $z_{\textrm{redmagic}}$. In the left plot, the full redMaGiC sample has a scatter of less than 0.02 and bias less than 0.01 over the entire redshift interval. 

\subsection{DES SN data}
In this analysis, we use data from the DES Supernova program (DES-SN) obtained with the Dark Energy Camera (DECam; \citealp{Flaugher15}) mounted on the 4 meter Blanco telescope at the Cerro Tololo Inter-American Observatory (CTIO). DES-SN operated over five seasons, taking observations in the optical \textit{griz} filters in ten 3 square degree fields at a cadence of $\sim7$ days. These images were preprocessed by the DES Data Management team (DESDM; \citealp{DESDM}). Next, transients were detected in these images using the Difference Imaging pipeline (DIFFIMG; \citealp{Kessler15}) by subtracting reference images from new observations. While scene modeling photometry (SMP; \citealp{Holtzman08, Astier14, Brout19a}) is planned for the entire DES-5YR photometric sample, we utilize DIFFIMG photometry that is calibrated at the $2\%$ level. This precision is sufficient for the purposes of this redMaGiC analysis, as calibration errors have the same effect on distances for spec-$z$ and photo-$z$, and here we only report differences between these two analyses. The transient sample is defined after restricting candidates to those with at least two detections at the same location on two separate nights in any band and that pass an automated artefact rejection algorithm (AUTOSCAN; \citealp{Goldstein15}). From these criteria, approximately 30,000 transients are identified, which include SNe, AGNs, and other transients and artefacts.

For each transient, a host galaxy is assigned using the directional light radius (DLR) method \citep{Sullivan06, Gupta16, Popovic20}. This host matching is performed with the depth-optimized coadds from \citet{Wiseman20}. The Photometric Supernova IDentification software (PSNID; \citealp{Sako11}) was run during survey operations on every active candidate, which fit the transient light-curve and provided preliminary classifications. This information informed the targeting for the host-galaxy follow-up spectroscopic program \citep{Smith&DAndrea20}.

The DES-5YR SN-like photometric sample is restricted to SNe with associated host-galaxy spectroscopic redshifts. These redshifts are obtained from the spectroscopic follow-up \citep{Smith&DAndrea20} program, primarily from the OzDES survey \citep{Yuan15, Childress17, Lidman20}. OzDES is a 100-night program using the 2dF+AAOmega spectrograph on the 3.9 meter Anglo-Australian Telescope. External redshift catalogues from the literature \citep[as cited in Table 1 of][]{Vincenzi21a} are used to supplement and optimize this redshift information. Following OzDES selection cuts and host associations, we have 5,049 galaxies with secure redshifts.

We further restrict the DES-5YR photometric sample (with no classifier applied) to SNe that have redMaGiC host galaxies. To associate host galaxies, we find all SNe with DLR-assigned host galaxy RA/Dec coordinates matching within one arcsecond to a redMaGiC galaxy from the Y6 run. 227 SNe, approximately 6$\%$ of the $\sim$3700 SNe fit by SALT2 in the 5YR sample, have redMaGiC host galaxies. The right side of Figure \ref{fig:RM_zbias} shows the redshift performance for the subset of redMaGiC galaxies with a SN match. The median redshift bias for the redMaGiC SNe is $\Delta z = 0.008$, and we find a redshift-dependent trend of $\sim0.006z$, which we show in Figure \ref{fig:RM_zbias}. We average the bias measured across 100 random samples of 227 galaxies drawn from the full redMaGiC distribution to obtain a mean bias of $0.001$ with a scatter of $0.001$, implying that the mean bias for the redMaGiC SNe is a $>5\sigma$ fluctuation from the full redMaGiC sample bias. As it is therefore possible that the subpopulation of redMaGiC SNe has an unmodeled selection effect, we propagate the $\sim0.006z$ bias as a systematic in Section \ref{sec:HDs}. The mean scatter for the subsample is 0.015, compared to 0.012 for the full sample, and 0.015 from the averaged same-size samples.

In addition to the DES-SN sample, we include an external spectroscopically confirmed low-redshift sample to anchor the Hubble Diagram. We use 182 SNe Ia from the Foundation Supernova Survey \citep{Foley18, Jones18}. We note that this low-redshift sample contains SNe from a range of host-galaxy types.

\subsection{Light-curve Fits, Distance Estimation, and Cosmological Parameter Recovery}\label{sec:2.3}

We use the SALT2 light-curve model \citep{Guy07, Guy10} to fit SN light-curves and standardize the SNIa brightnesses. This fitting is implemented in the SuperNova ANAlysis software (SNANA; \citealp{SNANA}) framework, based on the {\tt MINUIT} $\chi^2$ minimization algorithm to obtain best fit parameters and uncertainties. The fitted parameters are color $c$, stretch $x_{1}$, epoch of SN peak brightness $t_{0}$, and the overall amplitude $x_{0}$, with $m_{B} = -2.5\log_{10}(x_{0})$. The distance modulus, $\mu$, is estimated using the Tripp estimator \citep{Tripp, Astier06}:

\begin{equation}
    \mu= m_B + \alpha x_1 - \beta c - M_{z_i} +\delta \mu_{\textrm{bias}}
    \label{eq:Tripp}
\end{equation}
where $M_{z_i}$ is the distance offset in redshift bins $z_i$, $\alpha$, $\beta$ are coefficients parametrizing the relationship between stretch, color, and luminosity, and $\delta \mu_{\textrm{bias}}$ is the distance bias correction, which is described below. Rather than fitting for nuisance parameters ($\alpha$, $\beta$) in a global fit, we use values measured by the Dark Energy Survey \citep{Brout19b}; $\alpha=0.145$, $\beta=3.1$. 

Before fitting with SALT2, the data is further restricted to transients that have at least two bands with data satisfying max SNR $>4$. After fitting with SALT2, we apply the following selection requirements that are typical for a cosmology analysis, along with an additional color uncertainty requirement: 
\begin{itemize}
    \item fitted color $\left| c \right| < 0.3$
    \item fitted color uncertainty $\sigma_{c} < 0.2$
    \item fitted stretch $\left| x_1 \right| < 3.0$
    \item fitted stretch uncertainty $\sigma_{x_1} < 1.0$
    \item fitted $t_0$ uncertainty $\sigma_{t_0} < 2.0$ days
    \item Milky Way color excess $E(B-V) < 0.3$
\end{itemize}

Using the Beams with Bias Corrections (BBC; \citealp{Kessler17}) formalism, we bias correct our distance modulus values. The simulations used to determine bias corrections (biascor) are detailed in Section \ref{sec:sims}. Typical cosmological Ia analyses use higher dimensional bias corrections (BBC5D/BBC7D/BBC-BS20; \citealp{Popovic21}), with biases binned along SN properties, and use simulations with a $2\times2$ grid of $\alpha$ and $\beta$ (to interpolate for each fitted $\alpha$, $\beta$). However, to simplify this first redMaGiC analysis we use redshift-dependent (1D) bias corrections with no dependence on other parameters, and our simulations have fixed $\alpha$ and $\beta$. The BBC fit determines an intrinsic scatter term ($\sigma_\textrm{int}$) and $M_{z_i}$ in $N_z$ BBC redshift bins. From the SALT2- and BBC-fitted parameters, we compute a bias-corrected distance (Eq. \ref{eq:Tripp}) for each SN to use in the cosmology fit.

The uncertainty in $\mu$ is given by:
\begin{equation}
    \sigma_{\mu}^{2} = \sigma_{\textrm{int}}^{2} + \sigma_{\textnormal{Tripp}}^{2} + (\sigma_{\mu}^{\textrm{vpec}})^{2} + (\sigma_{\mu}^{z})^{2}
    \label{eq:sig_mu}
\end{equation}
where $\sigma_{\textrm{int}}$ is the intrinsic scatter needed to achieve reduced $\chi^2 ({\chi^2}/\textrm{degree of freedom})=1$ for the BBC fit, $\sigma_{\textrm{Tripp}}$ describes the uncertainty computed from fitted light-curve parameters and their covariances, $\sigma_{\mu}^{\textrm{vpec}}$ is the contribution from peculiar velocity uncertainty $\sigma_{\textrm{vpec}}$ with:
\begin{equation}
    \sigma_{\mu}^{\textrm{vpec}} = \left(\frac{5}{\ln(10)}\right) \frac{1+z}{z(1+z/2)} \left(\frac{\sigma_{\textrm{vpec}}}{c}\right)
    \label{eq:sig_muz}
\end{equation}
and $\sigma_{\mu}^{z}$ is the contribution from redshift uncertainty $\sigma_{z}$.

In Appendix A1, we explain why the redshift uncertainty ($\sigma_{z}$) is not included in Equation \ref{eq:sig_muz} as it has been previously in \citet{Kessler17}. Instead, an empirical method is used to account for the contribution of photo-$z$ uncertainties to the uncertainty in $\mu$ ($\sigma_{\mu}^{z}$) as explained in Section \ref{sec:redshift_color}. 

Following recent studies of systematic biases in SNIa cosmological analyses \citep{BHS21, Popovic21}, we fit for $w$ using a Gaussian prior on $\Omega_{\textrm{M}}$ of mean 0.311 and $\sigma=0.01$. We use the ``{\tt wfit}'' $\chi^{2}$ minimization program implemented in SNANA. 

\renewcommand{\arraystretch}{1.5}

\begin{table}[h!]
  \centering
\begin{tabular}{m{0.5\linewidth}|m{0.17\linewidth}|m{0.17\linewidth}}

Post-SALT2 fit cut & Number remaining & Number rejected\\
\hline \hline
SNe in redMaGiC galaxies fit by SALT2 & 224 & \\
\hline
SN color $\left| c \right| < 0.3$ & 185 & 39\\
\hline
SN color uncertainty $\sigma_{c} < 0.2$ & 184 & 1\\
\hline
SN stretch $\left| x_1 \right| < 3.0$ & 147 & 37\\
\hline
SN stretch uncertainty $\sigma_{x_1} < 1.0$ & 125 & 22\\
\hline
SN $t_0$ uncertainty $\sigma_{t_0} < 2.0$ & 125 & 0\\
\hline
Milky Way color excess $E(B-V) < 0.3$ & 125 & 0\\
\end{tabular}
\caption{Summary Table of Post-SALT2 fit cuts}
\label{tab:cuts}
\end{table}

\section{Simulations}\label{sec:sims}

To quantify the cosmological biases from using photometric redshifts, we use SNANA catalog level simulations to realistically represent the DES-5YR photometric SN sample and to analyze alongside the real data. We exclude core-collapse SN contamination from our simulations, as significant contamination is not expected, as discussed in Section \ref{sec:classification}. In addition, we generate large samples used by BBC for determining bias corrections \citep{Kessler17} to correct for known selection effects in our analysis.

\subsection{Baseline DES Simulation}
We use the SNANA software with the Pippin pipeline \citep{Pippin} to produce our simulations. Following the detailed simulations developed in \citet{Kessler19}, we generate realistic transient light-curves with several modifications. Briefly, the simulation consists of three major steps. First, a source SED is generated, and various astrophysical effects such as cosmological dimming, galactic extinction, lensing, and redshift effects are applied. The SED model is then integrated over the DES filters and observational noise is added using the DES observing conditions (PSF, sky noise, photometric zeropoints). Lastly, the detection efficiency and spectroscopic selection function of DES are implemented to select simulated events. To describe the SNIa brightness variation, we vary the SALT2 SED using the G10 intrinsic scatter model from \citet{Kessler13}. 

\citet{Vincenzi21a}, hereafter V21, makes several improvements to the DES-3YR simulations to replicate the DES-5YR photometric SN sample, which contains both core-collapse SNe and SNe Ia, but not other transients. We provide an overview of the important features utilized in our simulations. First, V21 uses a model of the host-galaxy spectroscopic redshift efficiency which is parametrized as a function of host-galaxy brightness, color, and year of discovery. V21 also improves on the host-galaxy library from \citet{Smith20} by compiling galaxy masses and Star Formation Rates (SFR) and accounting for differing SN rates in different types of galaxies. Simulating SN host galaxies based on published SN rates ensures that host-galaxy property dependencies are appropriately accounted for and that selection effects are modeled accurately across different galaxy types.

\subsection{Modifications to Baseline DES-5YR Simulation}\label{sec:3.2}
To properly simulate SNIa samples, the underlying distributions of SALT2 parameters $x_{1}$ and $c$ must accurately reproduce our observations. Previous analyses \citep{Scolnic18, DES3YR} have used the method in \citet{Scolnic16} to determine a migration matrix describing the impact of selection effects, noise, and intrinsic scatter on the underlying distributions. We use this methodology as implemented in the parent population fitting program from \citet{Popovic21} to account for the host-galaxy stellar mass relationship. These parent populations are described by an asymmetric Gaussian. In comparison to the population fits for the full DES-3YR spectroscopic sample, we find that the redMaGiC population parameters differ in color and stretch. When averaged over host-galaxy mass bins, the subset redMaGiC population is described by parameters as given in \autoref{tab:parentpops}. We provide the full parent population parameters for both the subset redMaGiC population and the full DES-3YR spectroscopic sample in Appendix A2. We highlight that the lower value of mean $x_1$ for the redMaGiC subsample is consistent with previous studies that show higher mass galaxies and lower specific star formation rates (sSFR) are correlated with lower $x_1$ values \citep{Childress14, Rigault20, Nicolas21}. 

We modify the V21 host-galaxy library and select only passive, bright galaxies to mimic the selection of redMaGiC galaxies. Selecting only passive galaxies as determined by $\textrm{log(sSFR)} < -11.5$ $\textrm{yr}^{-1}$ \citep{Sullivan06, Vincenzi21a}, we apply a cut on $r$ band magnitude in the host-galaxy redshift efficiency map, $m_r < 23.3$, and a cut on host-galaxy mass, $\textrm{logMass} > 10.5$, based on the distributions for the entire redMaGiC galaxy sample.

\begin{figure}
	\centering
	\includegraphics[scale=.35]{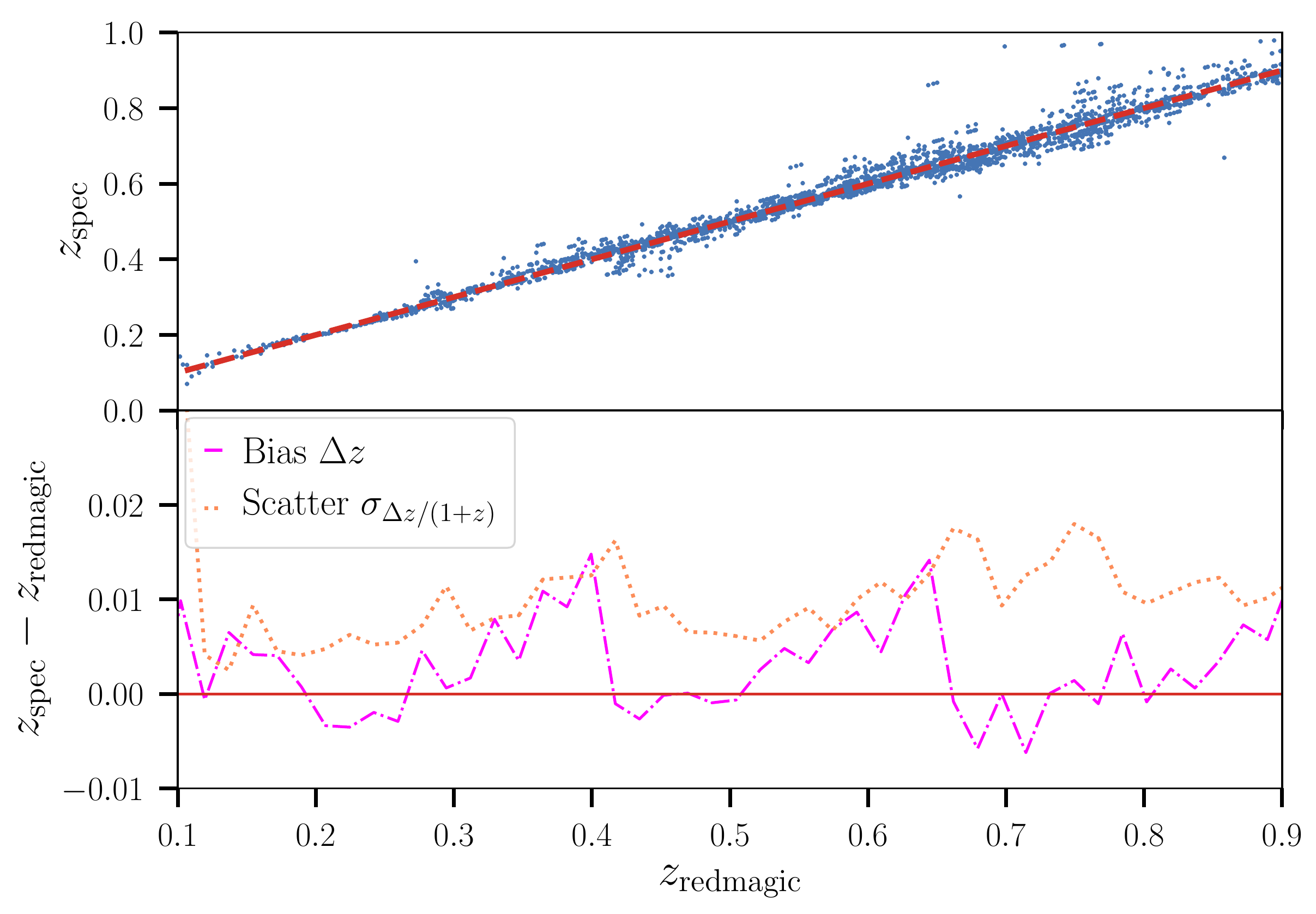}
	\caption{Redshift bias and scatter as defined in Figure 2, but for a random selection of 3,000 SNe from simulations.}
	\label{fig:sim_zbias}
\end{figure}

To accurately simulate the photo-$z$ biases and scatter in using photometric redshifts, we include a photo-$z$ for each galaxy in the host-galaxy library based on the bias and scatter from the redMaGiC catalog. For each galaxy in the modified V21 host-galaxy library, we find its closest match in redshift to the redMaGiC catalog, evaluate the bias $z_{\textrm{redmagic}} - z_{\textrm{spec}}$ for the redMaGiC galaxy, and add this bias to the host-galaxy true redshift value to determine the photo-$z$. While this galaxy matching would ideally be weighted by mass to prevent preferentially selecting lower mass galaxies, we find that the range of masses for redMaGiC host galaxies is narrow (further discussed in Section \ref{sec:disc}), circumventing this concern.  Figure \ref{fig:sim_zbias} shows that the photo-$z$ redshift bias and scatter are well reproduced with respect to the data. These photometric redshifts are propagated into the simulated data.

As discussed in Section \ref{sec:disc}, we measure the color-luminosity relation $\beta$ of 2.0 from the data and therefore simulate our redMaGiC-hosted SN sample with $\beta$ of 2.0 as well. A summary of the SN and host galaxy simulation properties is given in \autoref{tab:simsummary}.

\subsection{Comparison with Data}
\begin{figure*}
	\centering
	\includegraphics[scale=.35]{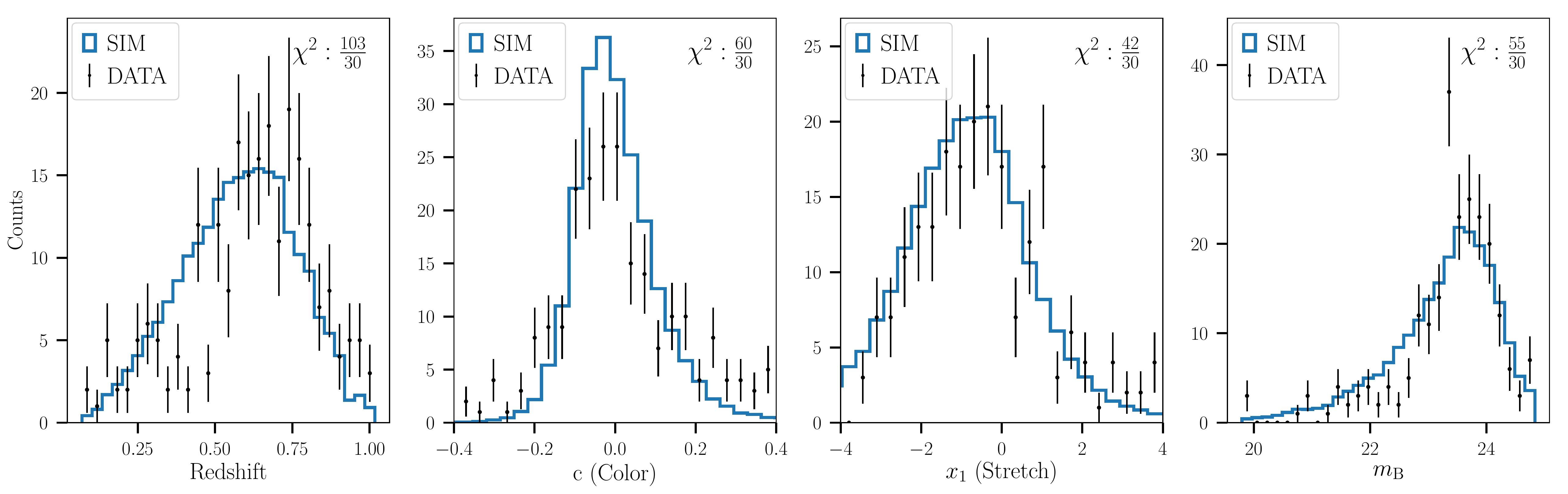}
	\caption{Distributions of redshift (\textbf{left}) and light-curve parameters: $c$ (\textbf{middle left}), $x_1$ (\textbf{middle right}), $m_B$ (\textbf{right}) for data and simulations. The data are represented by the black points, while simulations are represented by the blue histograms. Simulation histograms are normalized to the data. The reduced $\chi^2$ value is reported for each parameter. Covariances between parameters are not shown, as they are found to be small.}
	\label{fig:sim_data}
\end{figure*}

To validate the simulation, we normalize the total number of simulated SNe to the total number of observed SNe and find that our simulations reproduce the redshift distributions from the data well, as seen in Figure \ref{fig:sim_data}. We find that the light-curve parameter ($c$, $x_1$, $m_B$) distributions are also well reproduced. For redshift, we find reduced $\chi^2=$  \chisqredshift{}. For light-curve parameters $c$, $x_1$, and $m_B$, we find $\chi^2=$ \chisqcolor{}, \chisqstretch{}, and \chisqmb{} respectively. While the $\chi^2$ value for redshift is somewhat high due to the discrepant bins around $0.3<z<0.5$, the current DES-5YR analysis (Fig. 7 in V21) reports a comparable reduced $\chi^2$ value for the shallow fields of 65/22. There are likely some unmodeled selection and noise effects. We note that the redshift agreement could be improved if the redMaGiC algorithm is applied to the host-galaxy library, rather than the rough cuts utilized here, but this is beyond the scope of this work. Notably, the general shape agreement at higher redshift ($z > 0.55$) is of the greatest relevance, as it is where the distance bias correction impact is largest. For $c$, $x_1$, and $m_B$, our $\chi^2$ values are also comparable to V21, which finds values of 62/22, 29/21, and 37/19 respectively.

To compare the simulated host-galaxy photometric redshifts to data, we examine the mean redshift bias and scatter in the redshift range 0.1-0.9 as illustrated in Figures \ref{fig:RM_zbias} and \ref{fig:sim_zbias}. The combined high-density and high-luminosity redMaGiC catalog has mean redshift bias of 0.0005 and mean redshift-binned scatter of 0.0106 without any outlier cut. A random selection of 3,000 simulated host-galaxy redshifts has mean bias of 0.0003 and RMS scatter of 0.0099. Across redshift bins, the simulated redshifts have sub-percent bias and scatter of less than 0.02, accurately reproducing the data. 

\begin{table}[!thbp]
  \centering
\begin{tabular}{cc|c|c|c}
& & mean & $\sigma_{-}$ & $\sigma_{+}$ \\
\hline \hline
\multirow{2}{8em}{SNe with redMaGiC hosts} & $c$ & $-0.04$ & $0.04$ & 0.13 \\
& $x_1$ & $-0.86$ & 1.69 & 1.35 \\
\hline
\multirow{2}{8em}{DES-3YR sample} & $c$ & $-0.09$ & 0.02 & 0.16 \\
& $x_1$ & 0.15 & 1.01 & 0.66 \\
\end{tabular}
\caption{Parent Population Parameters, averaged over host-galaxy mass bins}
\label{tab:parentpops}
\end{table}

\begin{table}[!thbp]
  \centering
\begin{tabular}{cl|c}
\hline \hline
\multicolumn{2}{c|}{\textbf{SNIa Property}} & \\
\hline
& SED model & SALT2.JLA-B14\_LAMOPEN \\
\cline{2-3}
& SED variation & G10\\
\cline{2-3}
& $\alpha$ & 0.15\\
\cline{2-3}
& $\beta$ & 2.0\\
\cline{2-3}
& Parent Populations & see \autoref{tab:parentpops}\\
\hline \hline
\multicolumn{2}{c|}{\textbf{HOSTLIB (V21)}} & \\
\hline
& Host logMass & $>10.5$\\
\cline{2-3}
& Host $r$ mag & $<23.3$ \\
\cline{2-3}
& Host log(sSFR) & $<-11.5$ (passive) \\
\hline

\end{tabular}
\caption{Simulation Inputs and SN Properties}
\label{tab:simsummary}
\end{table}

\section{Analysis and Results}\label{sec:analysis}

We next discuss the results from our use of redMaGiC photometric redshifts for SNIa cosmology. In Section \ref{sec:classification}, we quantify and discuss the SNIa purity for supernovae in redMaGiC galaxies. In Section \ref{sec:results_sims}, we describe the performance of photo-$z$s in comparison to spec-$z$s for our simulations and the effects on cosmological parameters. In Section \ref{sec:results_data}, we apply our methods to data and compare the results with our simulations.

\subsection{Potential Core-collapse Contamination}\label{sec:classification}
Large elliptical galaxies such as LRGs are expected to contain mostly SNe Ia \citep{Hakobyan12}, with very low rates of core-collapse SNe. We quantify the potential core-collapse contamination in our redMaGiC subsample by examining photometric light-curve classification probabilities. The classification probabilities are obtained with the SuperNNova (SNN; \citealp{Moller20}) photometric SN classifier trained on the ``baseline" DES-like simulation presented in \citealt{Vincenzi21b}, which is generated from SEDs that include SALT2 and V19 core-collapse templates \citep{Vincenzi19}. SNN gives a probability ranging from 0 to 1.0, with 1.0 being most-likely-Ia. To consider potential contaminants, we define an unlikely-Ia SN as having probability of being a Type Ia (PIa) of $<0.5$. With conventional SALT2 cosmology cuts \citep{Betoule14}, color uncertainty cut as included in this analysis, and no classifier, the simulated DES-5YR photometric SN sample includes 8\% contamination \citep{Vincenzi21a, Moller22}. Of the 125 SNe in redMaGiC galaxies that pass post-SALT2 cuts, 4 are classified by SNN as unlikely-Ia ($\sim$ 3\%).

For comparison, we consider the SNN classification probabilities for the DES-5YR spectroscopically-confirmed SNIa sample, which serves as a ``truth'' set of SNe Ia. Of the 401 ``true'' SNe Ia, 3 are classified by SNN as non-Ia ($\sim$ 1\%). These fractions and percentages are shown in \autoref{tab:classification}. We also consider the sample of DES-5YR spectroscopically-confirmed non-Ia SNe and find that there is no overlap with the redMaGiC SN sample.

\renewcommand{\arraystretch}{1.3}

\begin{table}[h!]
\begin{tabular}{m{0.5\linewidth}|m{0.4\linewidth}}
\hline \hline
& Fraction (\%) of SNe classified by SNN as unlikely-Ia\\
\hline \hline
Simulated DES-5YR photometric SN sample with no classifier & 135/1680 (8\%)\\
\hline
SNe in redMaGiC galaxies & 4/125 ($\sim 3\%$)\\
\hline
DES-5YR spectroscopically-confirmed SNIa sample & 3/401 ($\sim 1\%$)\\
\hline
\end{tabular}
\caption{Potential Core-collapse Contamination}
\label{tab:classification}
\end{table}

Of the four events in the redMaGiC subsample classified as non-Ia by SNN, one is a spectroscopically confirmed Type Ia, and two are classified as Ia when SNN is trained instead on J17 templates \citep{Jones17a}. The baseline SNN model for DES in \citet{Moller22} also classifies one of the events as Ia. In particular, we note that in the Difference Imaging pipeline, the misclassified spectroscopically confirmed SNIa has inaccurately subtracted template images in the $g$ band. Notably, photometric classifiers do not have 100\% efficiency and can miss true SNe Ia.  We confirm the claim that redMaGiC galaxies have very low rates of core-collapse SNe.

\subsection{Results from Simulations}\label{sec:results_sims}
We utilize our simulations as described in Section \ref{sec:sims} to characterize the systematic effects of using photometric redshifts to measure $w$. Within the SNANA framework, we measure cosmological parameters using both the spectroscopic redshift and the redMaGiC host-galaxy photo-$z$ as described in Section \ref{sec:3.2}.

With the resulting light-curves, we perform the analysis detailed in Section \ref{sec:2.3} and obtain $\mu$ for each event. The true distance in each biascor is computed from the measured redshift, either the spec-$z$ or the redMaGiC host-galaxy photo-$z$. While the dispersion of the Hubble residuals will be larger as a result of using 1D (redshift) bias corrections instead of 5D ($z, x_1, c, \alpha,\beta $), we clarify that there are no additional biases introduced, because our simulations accurately model the data for each redshift case and therefore produce the appropriate bias correction. We study the impact of using incorrect bias corrections due to potential mismodeling of redshift bias and scatter in Section \ref{sec:HDs}. The bias corrections for the anchoring low-redshift sample are computed separately. Although we simulate the DES sample with $\beta=2.0$, for this study in the BBC step we fix $\beta=3.1$ as described in Section \ref{sec:2.3} due to the inclusion of the low-$z$ anchor. We again clarify that any biases introduced due to the use of samples with different $\beta$ will be the same when using spectroscopic and photometric redshifts and therefore will not contribute to the $\Delta w$ values reported here.
 
\subsubsection{Redshift Contribution to Distance Modulus Uncertainty}\label{sec:redshift_color}

To estimate the contribution to the distance modulus uncertainty from the photometric redshift uncertainty, we examine the difference in Hubble residual scatter between spec-$z$s and photo-$z$s using simulations. To avoid statistical limitations from the redMaGiC sample size, we enlarge the host galaxy library by applying the same photo-$z$ bias as described in Section \ref{sec:3.2} to the V21 host-galaxy library without the redMaGiC-like galaxy cuts. As a crosscheck, we also simulate a second set of photo-$z$ using a simple RMS map where the host-galaxy photo-$z$ scatter is described as a function of redshift (as seen in Figure \ref{fig:RM_zbias}). We find that for the DES redshift range in both sets of simulations the distance modulus uncertainty contribution from the use of photo-$z$s is $\sim0.06$ mag, which is small compared to the RMS from the distance measurement uncertainty (using the spec-$z$) of $\sim0.18$ mag or higher. For this work, we therefore neglect this contribution to the uncertainty in $\mu$ and note future methods for treating redshift uncertainties in Section \ref{sec:disc}.

\subsubsection{Hubble Diagrams and Cosmological Parameters}\label{sec:HDs}

\begin{figure*}
	\centering
	\includegraphics[scale=.55]{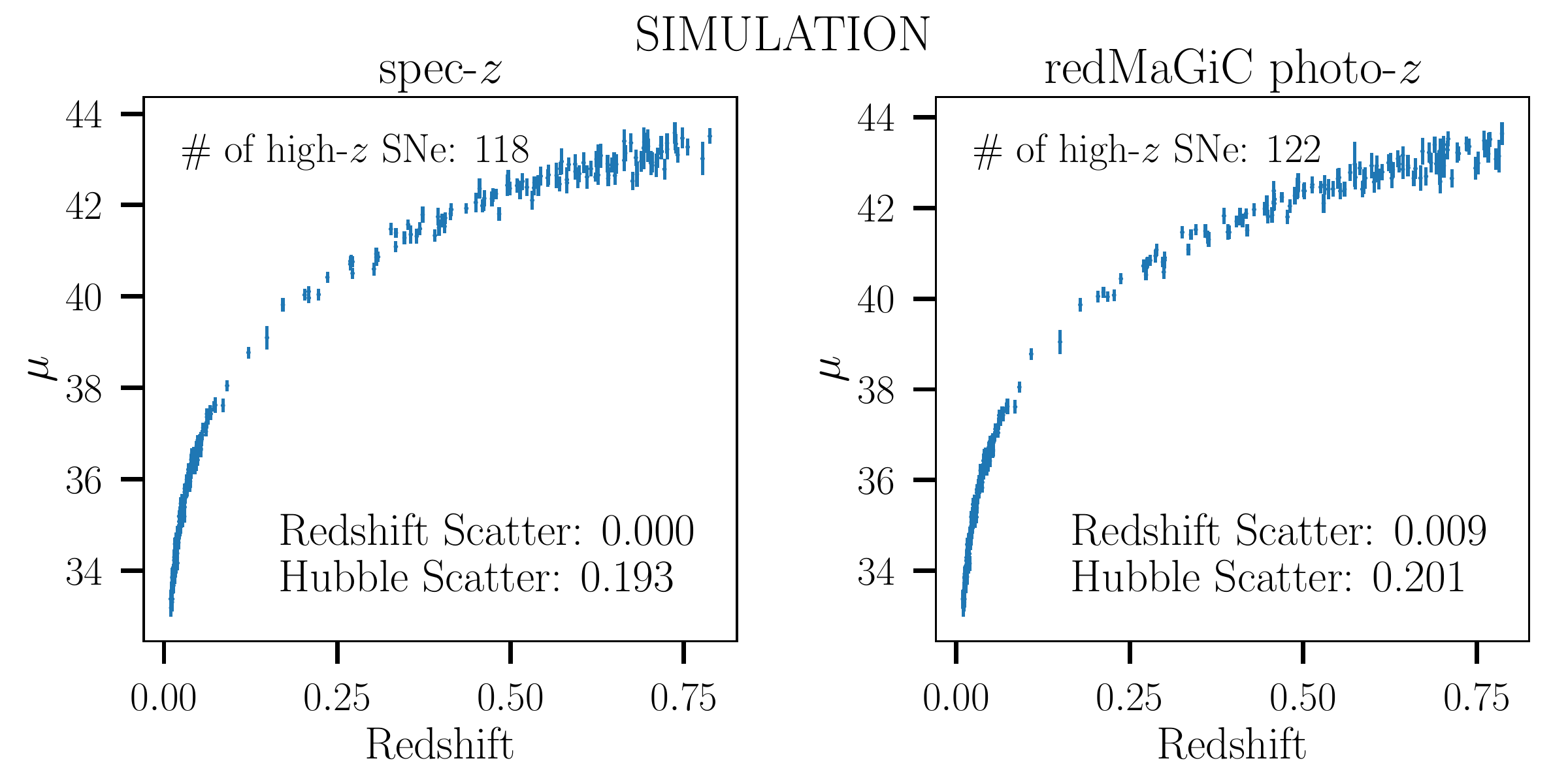}
	\caption{Hubble diagram from analyzing simulation for: spec-$z$ from host galaxy (\textbf{left}), redMaGiC photo-$z$s from the host galaxy (\textbf{right}). The number of high-$z$ SNe is shown on each panel, as well as the redshift scatter $\sigmadeltaz$ and Hubble scatter.}
	\label{fig:HDs_sim}
\end{figure*}

Using the spectroscopic and photometric host-galaxy redshifts, we analyze the simulated data to produce Hubble diagrams with $\sim 120$ simulated SNe, shown in Figure \ref{fig:HDs_sim}. The redshift scatter for spectroscopic redshift is by definition 0.00. We obtain a scatter of \ssigzphot{} for the redMaGiC photo-$z$ and compute the Hubble scatter (i.e., the scatter of the Hubble residuals) using a robust measure of the standard deviation, defined as $1.48 \times \textrm{MAD}$. The Hubble scatter for the spec-$z$ case is \shubblespec{}, while for the redMaGiC photo-$z$, the Hubble scatter is \shubblephot{}. As expected, both redshift and Hubble scatter are larger for redMaGiC photo-$z$ than spec-$z$ (summarized in \autoref{tab:scatter}). 

In \autoref{tab:deltaw}, we show $\Delta w$,
\begin{equation}
    \Delta w = w_\textrm{spec} - w_\textrm{phot}
    \label{eq:deltawdef}
\end{equation}
the difference in $w$ between using the spectroscopic redshift and the photo-$z$ method. This difference is averaged over 150 statistically independent simulations to obtain an average $\Delta w$. We find 
\begin{equation}
    \Delta w = \sdeltawone{} \pm \sdeltawerrone{}
    \label{eq:deltaw}
\end{equation}
with standard deviation \sdeltawstdone{}. 

Next, we examine systematic variations in which the simulated photo-$z$ bias or scatter does not match the data. To improve our $w$-bias estimate from systematics, we first evaluate the effect of an exaggerated shift and then assume a linear scale for a more realistic shift. For the photo-$z$ scatter systematic test, a biascor is generated with an analytic description of the host photo-$z$, where $z_{\textrm{phot}} - z_{\textrm{spec}}$ is drawn from a Gaussian of width $\sigma=0.03(1+z)$. We measure $\frac{d(\Delta w)}{d(\Delta \textrm{scatter})}=\frac{0.0107}{0.03}$ and estimate a realistic systematic of $0.0054$ given scatter of 0.015. For the photo-$z$ bias systematic test as mentioned in Section \ref{sec:data}, we measure the change in $w$ with respect to change in bias. A biascor is generated with the host photo-$z$ containing an additional bias of 0.006$z$. Applying the bias correction on the data, we measure $\frac{d(\Delta w)}{d(\Delta \textrm{bias})}=\frac{0.0171}{0.006}$, and therefore a potential systematic bias of $0.0171$.

Without utilizing the actual spectroscopic redshifts of our sample of 227 SNe to ascertain a potential systematic bias, we also consider a realistic test of potential calibration shifts of the photometric redshifts as used in current large-scale structure cosmological analyses. We generate a biascor with $z$-dependent bias as determined in the 2-parameter fit calibration of the redMaGiC galaxy sample using clustering redshifts, or `cross-correlation redshifts,' as used for DES weak lensing and galaxy clustering studies. \citet{Cawthon20} present the best fit shift and stretch parameters that are applied to the redMaGiC photometric redshift distributions to better match the redshifts determined via angular cross-correlation of the redMaGiC sample with spectroscopic galaxy samples. We parameterize this shift (given as $\Delta z$ in \citealt{Cawthon20}, Table 7) as a function of redshift using cubic spline interpolation and add it to the photo-$z$ biascor simulated as described in Section \ref{sec:3.2}. We find $\Delta w = \sdeltawthree{} \pm \sdeltawerrthree{}$ with standard deviation \sdeltawstdthree{}. These $\Delta w$ values are also shown in Table \ref{tab:deltaw}. While each of these biascor tests results in larger $\Delta w$, they remain consistent within the standard deviations. This indicates that the use of redMaGiC photo-$z$ is robust to potential mismodeling in our biascor simulations.

We also test the impact of having removed the $\sigma_z$ term from Equation \ref{eq:sig_muz} as described in Section \ref{sec:2.3} and Appendix A1. We find $\Delta w = 0.0098 \pm 0.0029$ with standard deviation 0.0361. The mean value of $\Delta w$ is not strongly impacted, but the uncertainties are increased as expected due to the overestimated redshift error when $\sigma_z$ is included.

\begin{table}[t]
  \centering
\begin{tabular}{m{0.15\linewidth}|m{0.16\linewidth} m{0.16\linewidth}| m{0.16\linewidth} m{0.16\linewidth}}
 & \multicolumn{2}{c}{Simulation} & \multicolumn{2}{c}{Data} \\
\toprule
           Methods & Redshift Scatter $\sigmadeltaz$ &  Hubble Scatter &  Redshift Scatter $\sigmadeltaz$ &  Hubble Scatter \\
\hline \hline
           spec-$z$ & \ssigzspec{} & \shubblespec{} & \dsigzspec{} & \dhubblespec{} \\
           \hline
           redMaGiC photo-$z$ & \ssigzphot{} & \shubblephot{} &   \dsigzphot{} &  \dhubblephot{} \\
           \hline
\end{tabular}
\caption{Redshift and Hubble Scatter for simulations and data}
\label{tab:scatter}
\end{table}

\begin{table}[t]
  \centering
\begin{tabular}{m{0.15\linewidth}|m{0.09\linewidth} m{0.09\linewidth}m{0.09\linewidth}| m{0.09\linewidth} m{0.09\linewidth}m{0.09\linewidth}}
 & \multicolumn{3}{c}{Simulation} & \multicolumn{3}{c}{Data} \\
\toprule
           Methods &  $\alpha$ &  $\beta$ &  $\sigma_{\textrm{int}}$ & $\alpha$ &  $\beta$ & $\sigma_{\textrm{int}}$ \\
\hline \hline
           spec-$z$ & 0.14 & 3.1 & \sspecsigint{} & 0.14 & 3.1 & \dspecsigint \\
           \hline
           redMaGiC photo-$z$ & 0.14 & 3.1 & \sphotsigint & 0.14 & 3.1 & \dphotsigint\\
           \hline
\end{tabular}
\caption{BBC nuisance parameters for simulations and data. $\alpha$ and $\beta$ are fixed, while $\sigma_{\textrm{int}}$ is floated.}
\label{tab:bbc}
\end{table}

\begin{figure*}
	\centering
	\includegraphics[scale=.55]{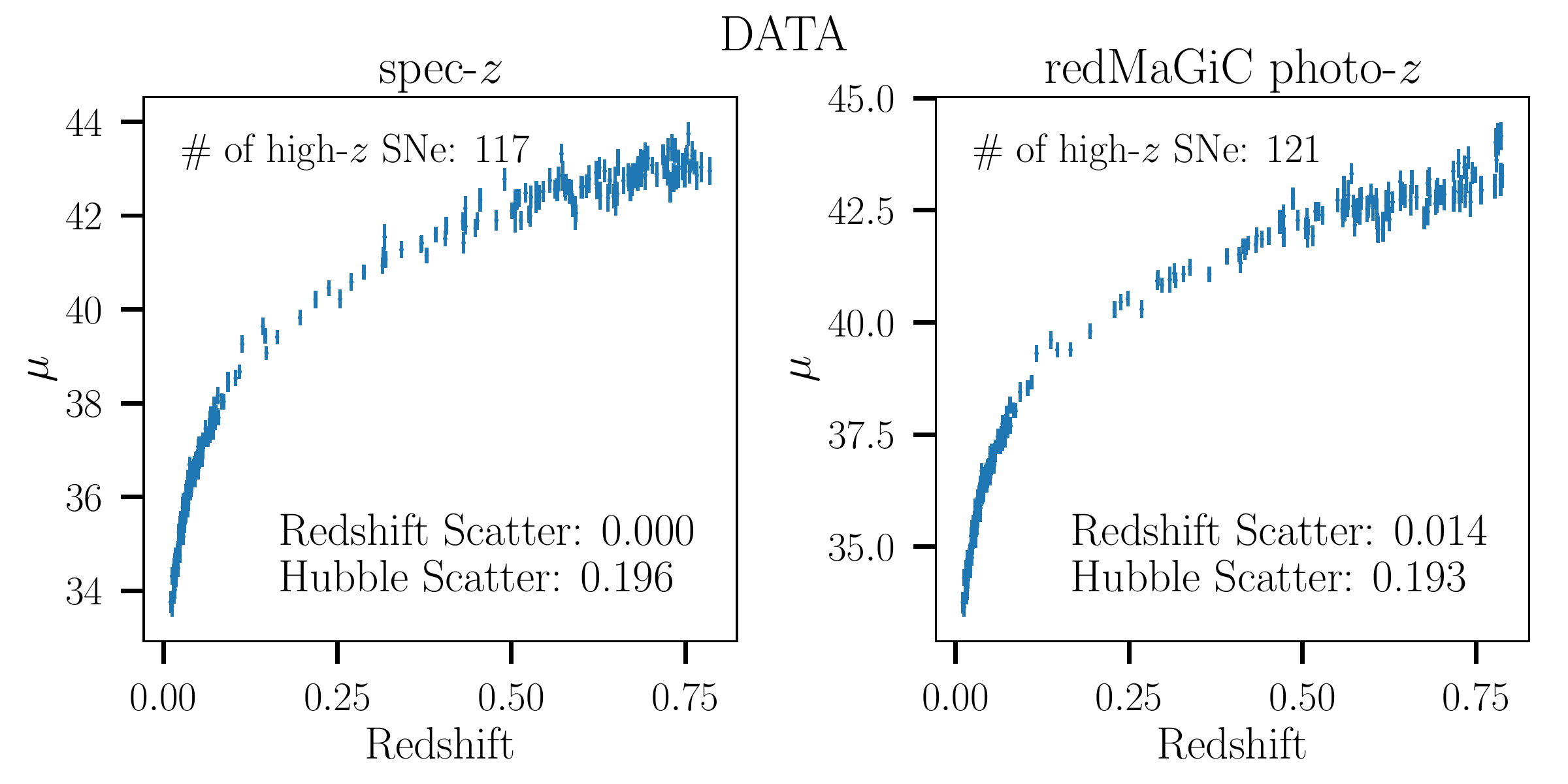}
	\caption{Hubble diagrams for redshift methods as described in Figure \ref{fig:HDs_sim}, but with data. The number of high-$z$ SNe, as well as the redshift scatter $\sigmadeltaz$ and Hubble scatter are given for each redshift method.}
	\label{fig:HDs_data}
\end{figure*}

\begin{table*}[t]
  \centering
\begin{tabular}{m{0.2\linewidth}|m{0.11\linewidth} m{0.1\linewidth} m{0.1\linewidth}| m{0.11\linewidth}m{0.15\linewidth}}
 & \multicolumn{3}{c}{Simulation} & \multicolumn{2}{c}{Data} \\
\toprule
           Methods &  $\Delta w$ &  $\Delta w$ Error &  $\Delta w$ STD & $\Delta w$ & $w$ Uncertainty\\
\hline \hline
           spec-$z$ & 0.00 & 0.00 & 0.00 & 0.00 & \dwuncertzero{} \\
           \hline
           redMaGiC photo-$z$ & \sdeltawone{} & \sdeltawerrone{} & \sdeltawstdone{} & \ddeltawone{} & \dwuncertone{} \\
           \hline
           RM photo-$z$ with additional $z$-dependent bias biascor & \sdeltawthree{} & \sdeltawerrthree{} & \sdeltawstdthree{} & \ddeltawthree{} & \dwuncertthree{}\\
           \hline
\end{tabular}
\caption{Difference in $w$ ($\Delta w$) between photo-$z$ and spec-$z$. The estimated uncertainty in $w$ for the spectroscopic case is \dwuncertzero{} for the data and on average \swuncertzero{} for the simulations.}
\label{tab:deltaw}
\end{table*}

\subsection{Results from Data}\label{sec:results_data}
The same methods used on simulations are applied to our data. For bias corrections, we generate biascor simulations as described in Section \ref{sec:sims}. Notably, the measured Hubble scatter when using the full redshift range is larger than predicted from the simulations (0.271 mag vs.\ 0.192 mag). This discrepancy is due to scatter at $z>0.8$, which is 0.451 mag for data and 0.238 mag for simulations. This indicates that we are not adequately modeling the SN light-curves at higher redshifts. Therefore, we limit the redshift range of our data and simulation samples to $z<0.8$, the upper limit of the redshift range for the DES-3YR spectroscopic sample. We obtain the Hubble diagrams shown in Figure \ref{fig:HDs_data} and the $\Delta w$ for data shown in \autoref{tab:deltaw}. For redMaGiC photo-$z$, we obtain redshift scatter of \dsigzphot{} and Hubble scatter of \dhubblephot{}. These figures are summarized in \autoref{tab:scatter}. We measure intrinsic scatter $\sigma_{\textrm{int}}=$ \dspecsigint{} for the data in comparison to \sspecsigint{} for our simulations. Similarly, we measure $\sigma_{\textrm{int}}=$ \dphotsigint{} in comparison to \sphotsigint{} for the photo-$z$ case. These floated BBC parameters, along with the fixed values for $\alpha$ and $\beta$ are provided in Table \ref{tab:bbc}.

When using redMaGiC photo-$z$, we obtain 
\begin{equation}
    \Delta w= \ddeltawone{}
\end{equation}
compared to the spectroscopic case. This result is consistent with our expectations from simulations (\sdeltawone{} $\pm$ \sdeltawerrone{} with standard deviation \sdeltawstdone) and is significantly smaller than the data $w$ uncertainty using spec-$z$ (\dwuncertzero{}), indicating that systematic uncertainties from using redMaGiC photo-$z$ are subdominant to our overall $w$ uncertainty. In summary, and as expected from simulations, replacing spectroscopic redshifts with photometric redshifts has a negligible impact on the width of the cosmological posteriors. Moreover, the systematic shift in the posterior is negligible compared to the width of the posterior.

\section{Discussion}\label{sec:disc}

\begin{figure}[htb]
	\centering
	\includegraphics[scale=.5]{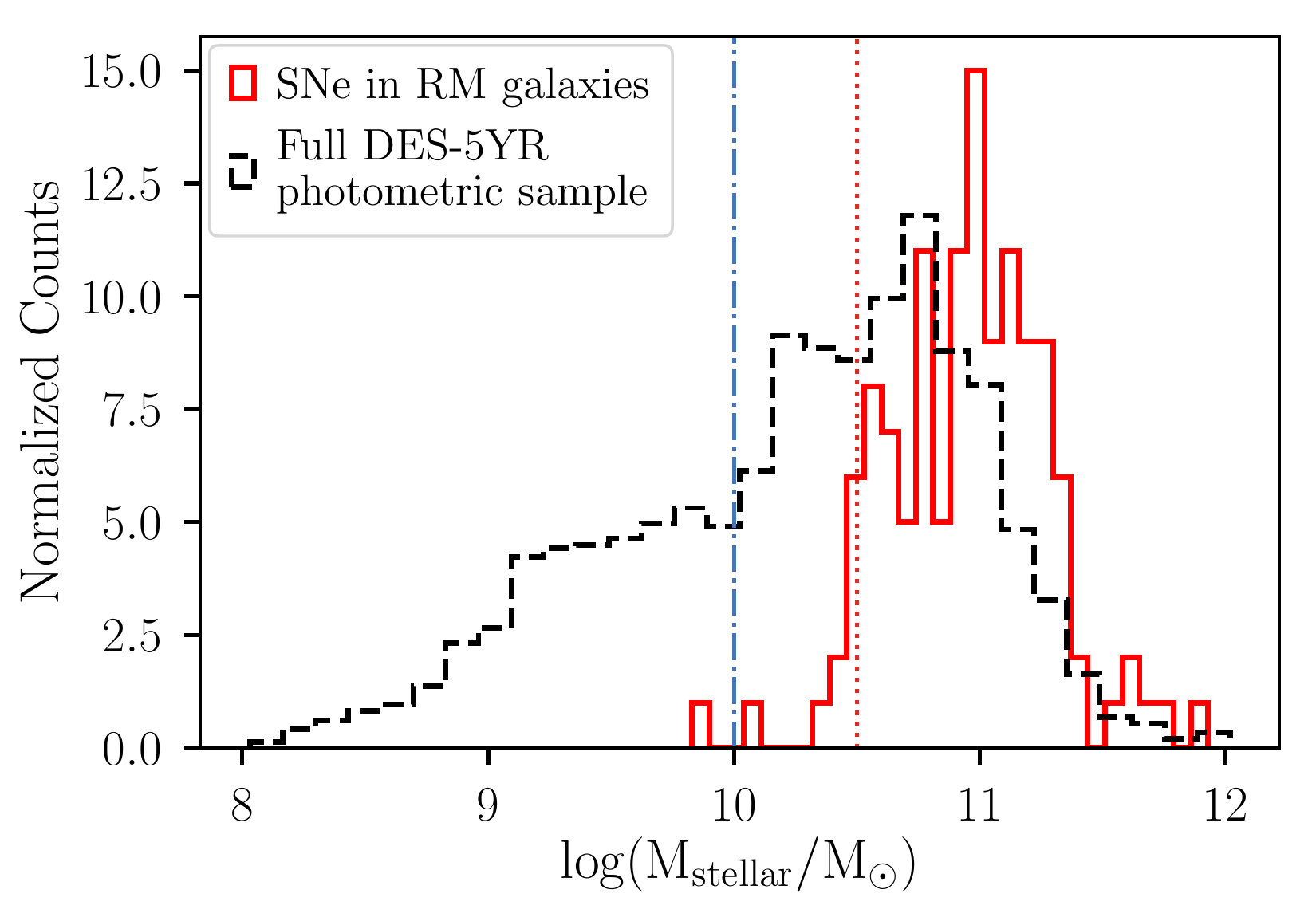}
	\caption{LogMass distribution for the full DES-5YR photometric sample in dashed black and for the subsample of SNe with redMaGiC host galaxies in solid red. The full DES-5YR sample is normalized to the redMaGiC subsample. The red dotted line at logMass of 10.5 indicates the cut made for the simulated host-galaxy library. The blue dash-dotted line indicates the cutoff typically applied to measure the mass step.}
	\label{fig:logmass}
\end{figure}

There are three advantages to a redMaGiC-only SN analysis: i) lower probability of of having non-Ia SNe, ii) ability to use host-galaxy photo-$z$, and iii) robustness to SN host-galaxy correlations. The first two of these have been covered extensively in this paper, so we here address the third. We note that of our redMaGiC SN subsample, only 1 out of 125 SNe has logMass $< 10$. In Figure \ref{fig:logmass}, we show the logMass distributions for the full DES-5YR photometric sample and for the redMaGiC SN subsample \citep{Smith20, Wiseman20}. About $90\%$ of our redMaGiC sample has $10.5<\textrm{logMass}<11.9$, a much narrower range compared to the full DES5YR photometric sample that spans $8 < \textrm{logMass} < 12$. Due to the range of host-galaxy masses for the redMaGiC subsample, it is impossible to measure the mass step, which is defined as the difference in intrinsic luminosity (after correction) between high mass galaxies (logMass $>10$) and low mass galaxies \citep{Smith20}. We expect that cosmological analyses using this subset of SNe Ia will be more robust to host-galaxy stellar mass dependencies, where here we use stellar mass as a proxy for other host-galaxy properties. Future studies may also find it worthwhile to consider the homogeneity of other host-galaxy properties such as star formation rate or metallicity.

When $\beta$ is floated in the BBC fit, we find $\beta=2.068\pm0.210$ for the redMaGiC sample, which is significantly smaller than $\beta=3.178\pm0.139$ for the DES-3YR spectroscopic sample (where $\beta$ or $\textrm{R}_\textrm{B}=\textrm{R}_\textrm{V}+1$). Interestingly, \citet{Meldorf21} find that the $\textrm{R}_\textrm{V}$ distribution for redMaGiC galaxies is at the lower end of $\textrm{R}_\textrm{V}$ range for the DES host-galaxy distribution ($\overline{\textrm{R}_\textrm{V}}=1.54\pm0.02$), whereas for the full distribution it is $\textrm{R}_\textrm{V}=2.61\pm0.07$. This finding supports the prediction from \citet{BS20} that the Hubble scatter vs. color relation can be explained by differing dust properties from different host-galaxy populations, as it shows a direct link between the low $\beta$ found for a particular subset of galaxies and a low $\textrm{R}_\textrm{V}$ predicted for this same set. \citet{Sullivan10} found that SNe with low-sSFR host galaxies have lower $\beta$ than SNe with high-sSFR hosts, and consistently, \citet{Kelsey20} found lower $\beta$ in high mass/redder rest-frame U-R galaxies than in low mass/bluer U-R galaxies. This finding indicates that our redMaGiC subsample of SNe is less sensitive to color. 

\subsection{Future Prospects}\label{sec:future}
As described in \citet{Kessler10a}, it is possible to use the SALT2 framework to fit a photometric redshift from the light-curve simultaneously with $x_{1}, c, m_{B}, t_{0}$. To improve photo-$z$ precision and reduce outliers, the host-galaxy photo-$z$ can be used as a prior for this SN light-curve fit. This 5-parameter fit is the technically correct method to account for redshift uncertainties so that they can be propagated to other SALT2 fitted parameters and covariances. However, our attempts to implement these methods presented problems at high-redshift, because only two passbands ($i$ and $z$) are within the SALT2 model range, which poorly constrains the SN color and redshift. In addition, the SALT2 model requires interpolations that result in occasional discontinuities in the model derivatives with respect to redshift, which can lead to pathological behavior in {\tt MINUIT}. Further investigation and resolution of these issues is beyond the scope of this paper, but should be considered for future work in photometric SN cosmology. An alternative method is to measure the redshift contribution to the distance modulus uncertainty empirically, as detailed in Section \ref{sec:redshift_color}. A further area of follow-up that will be required is improved modeling and better understanding of the Hubble scatter discrepancy between data and simulations at high redshifts ($z>0.8$) as explained in Section \ref{sec:results_data}.

SNIa cosmology with next generation surveys in the era of LSST and Roman will require methods such as the one presented here to make full use of the photometric data without the constraints imposed by the limits of spectroscopy. To make a simple forecast for LSST, we follow the simulations from the TiDES Collaboration \citep{Swann19, Frohmaier21}\footnote{\url{https://docushare.lsst.org/docushare/dsweb/Get/Document-37640/Frohmeier_TiDES.pdf}} using the Baseline 1.7 OpSim run. We assume that 6\% of the LSST SNe Ia discovered are in LRGs and that the redMaGiC photo-$z$ resolution is $\sigma_z=0.02$. In total, there are $\sim$2.4 million SNe Ia with two points of SNR $>$ 5 that pass light-curve quality cuts, of which 6\% is 144 thousand SNe. Assuming a low-$z$ sample of 2400 SNe to anchor the Hubble Diagram (LSST-DESC Science Requirements Document; \citealt{LSSTSRD}) and a prior on $\Omega_M$ of $0.311\pm0.01$, we recover an uncertainty on $w$ of $0.014$. This is $2\times$ smaller than the statistical constraint from the current sample of Pantheon SNe \citep{Scolnic18}. 

\citet{Mitra21} provide a quantitative requirement on photo-$z$ systematics for LSST and Roman and conclude that the redshift systematic from LSST color matched nearest neighbors (CMNN) photo-$z$ estimates must be reduced by an order of magnitude for unbiased SNIa cosmology. However, they do not propagate photo-$z$ biases through the light-curve fits, resulting in more stringent redshift requirements, as they do not account for the redshift-color correction noted in Appendix A1. Further studies and deeper understanding of the systematic uncertainties associated with photometric redshifts will be required to fully utilize the statistical power of such datasets. Appropriate treatment of photometric redshifts will require the inclusion of redshift in bias corrections (5D or 7D), rather than the 1D corrections used in this study. An alternative method is also presented in the zBEAMS hierarchical Bayesian formalism by \citet{Roberts17}. A Bayesian approach may improve the treatment of redshift errors and properly account for the redshift-color self-correction in data discussed in Appendix A1.

With increasing numbers of SN observations, it will be possible to use a low redshift sample that also contains only SNe in LRGs, providing us with a full sample of SNe Ia exclusively in red galaxies. However, the impacts of systematic redshift errors are much stronger at low redshift \citep{Wojtak15, Davis19}, and the color-redshift relation that reduces the impact of biases in the DES redshift range may not be sufficient for unbiased use of photometric redshifts at low-$z$. Further studies will also be required to determine the effects of redshift biases on constraining a time-varying equation-of-state. In the nearer future, with the Year 6 (Y6) analysis from DES and redMaGiC run on the deep fields, the number of SNe found to be hosted in red galaxies will increase. It may also be possible to investigate alternative methods of selecting LRGs in order to create a larger host-galaxy sample. Another extension that will require further investigation would be whether other galaxy types besides LRGs can be similarly used to restrict the host-galaxy type and reduce the photometric redshift bias. Lastly, it may be possible to include information about the distribution of redshifts beyond individual redshifts as is standard practice in other cosmology analyses such as the 3$\times$2 pt probes. The application of information traditionally used for other cosmology probes, such as the redMaGiC catalogs, to SNIa cosmology is a largely unexplored area with tremendous potential.

\section{Conclusion}\label{sec:conc}

Using the DES-5YR photometric sample, we present a first proof-of-concept that SNe Ia in redMaGiC hosts are a promising new avenue for SN cosmology. We show that restricting SNe to those with redMaGiC host galaxies serves as a useful cross-check for photometric classification, as they preferentially host only Type Ia SNe: $\sim 3\%$ of redMaGiC SNe are photometrically classified as non-Ia, compared to 8\% of a simulated DES-5YR photometric SN sample with no classifier.

We further present our cosmological parameter results and biases from using redMaGiC host-galaxy photo-$z$. Using redMaGiC photo-$z$ results in biases in $w$ of $\sim0.01$ when run on data, which is consistent with expectations from simulations. The Hubble scatter from data is \dhubblephot{} for redMaGiC photo-$z$, which is consistent with \shubblephot{} obtained from simulations. Our findings indicate that redMaGiC photo-$z$ can be used in a relatively unbiased manner with respect to spectroscopic redshifts. Lastly, we describe related extensions and potential future work using other sources of host-galaxy redshift information. This work lays the essential groundwork for future development of the use of photometric redshifts for SNIa cosmology in time-domain surveys, particularly for LSST and surveys with the Roman Space Telescope.

\section{Acknowledgements}
DS is supported by DOE grant DE-SC0010007 and the David and Lucile Packard Foundation. DS is supported in part by NASA under Contract No. NNG17PX03C issued through the WFIRST Science Investigation Teams Programme. Eduardo Rozo is supported by NSF grant 2009401. Eduardo Rozo is further supported by DOE grant DE-SC0009913, and by a Cottrell Scholar award. LK thanks the UKRI Future Leaders Fellowship for support through the grant MR/T01881X/1. LG acknowledges financial support from the Spanish Ministry of Science and Innovation (MCIN) under the 2019 Ram\'on y Cajal program RYC2019-027683-I and from the Spanish MCIN project HOSTFLOWS PID2020-115253GA-I00.

This paper has gone through internal review by the DES collaboration. Funding for the DES Projects has been provided by the U.S. Department of Energy, the U.S. National Science Foundation, the Ministry of Science and Education of Spain, the Science and Technology Facilities Council of the United Kingdom, the Higher Education Funding Council for England, the National Center for Supercomputing Applications at the University of Illinois at Urbana-Champaign, the Kavli Institute of Cosmological Physics at the University of Chicago, the Center for Cosmology and Astro-Particle Physics at the Ohio State University, the Mitchell Institute for Fundamental Physics and Astronomy at Texas A\&M University, Financiadora de Estudos e Projetos, Fundação Carlos Chagas Filho de Amparo à Pesquisa do Estado do Rio de Janeiro, Conselho Nacional de Desenvolvimento Científico e Tecnológico and the Ministério da Ciência, Tecnologia e Inovação, the Deutsche Forschungsgemeinschaft and the Collaborating Institutions in the Dark Energy Survey.

The Collaborating Institutions are Argonne National Laboratory, the University of California at Santa Cruz, the University of Cambridge, Centro de Investigaciones Energéticas, Medioambientales y Tecnológicas-Madrid, the University of Chicago, University College London, the DES-Brazil Consortium, the University of Edinburgh, the Eidgenössische Technische Hochschule (ETH) Zürich, Fermi National Accelerator Laboratory, the University of Illinois at Urbana-Champaign, the Institut de Ciències de l’Espai (IEEC/CSIC), the Institut de Física d’Altes Energies, Lawrence Berkeley National Laboratory, the Ludwig-Maximilians Universität München and the associated Excellence Cluster Universe, the University of Michigan, NFS’s NOIRLab, the University of Nottingham, The Ohio State University, the University of Pennsylvania, the University of Portsmouth, SLAC National Accelerator Laboratory, Stanford University, the University of Sussex, Texas A\&M University, and the OzDES Membership Consortium.

Based in part on observations at Cerro Tololo Inter-American Observatory at NSF’s NOIRLab (NOIRLab Prop. ID 2012B-0001; PI: J. Frieman), which is managed by the Association of Universities for Research in Astronomy (AURA) under a cooperative agreement with the National Science Foundation. 

Based in part on data acquired at the Anglo-Australian Telescope, under program A/2013B/012. We acknowledge the traditional owners of the land on which the AAT stands, the Gamilaraay people, and pay our respects to elders past and present.

The DES data management system is supported by the National Science Foundation under Grant Numbers AST-1138766 and AST-1536171. The DES participants from Spanish institutions are partially supported by MICINN under grants ESP2017-89838, PGC2018-094773, PGC2018-102021, SEV-2016-0588, SEV-2016-0597, and MDM-2015-0509, some of which include ERDF funds from the European Union. IFAE is partially funded by the CERCA program of the Generalitat de Catalunya. Research leading to these results has received funding from the European Research Council under the European Union’s Seventh Framework Program (FP7/2007-2013) including ERC grant agreements 240672, 291329, and 306478. We acknowledge support from the Brazilian Instituto Nacional de Ciência e Tecnologia (INCT) do e-Universo (CNPq grant 465376/2014-2).

This work was completed in part with resources provided by the University of Chicago’s Research Computing Center.


\newpage


\appendix

\subsection{A1. Redshift Uncertainties and Redshift Color Dependencies}
\label{sec:A1}
\begin{figure*}[h!]
	\centering
	\includegraphics[scale=0.45]{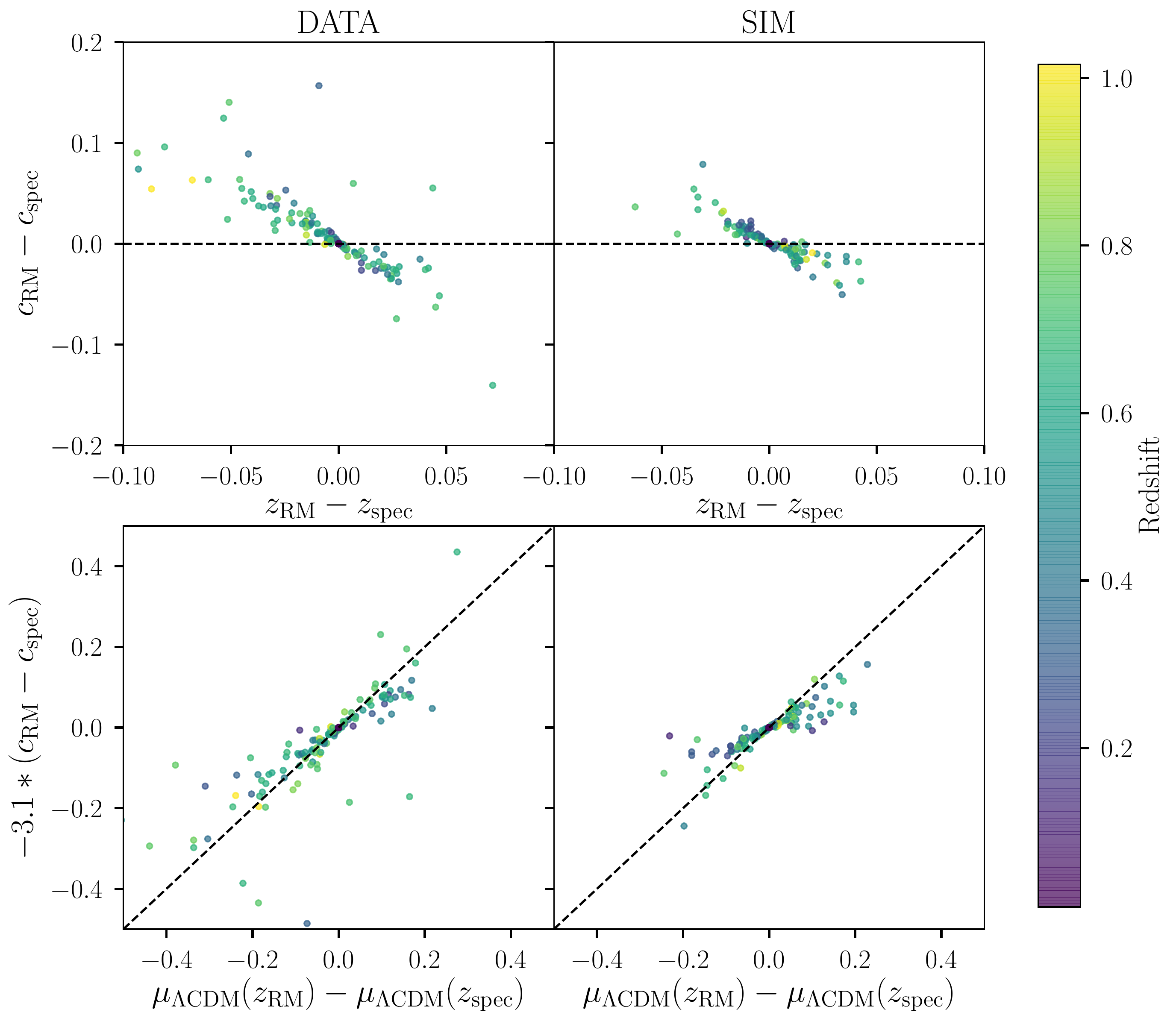}
	\caption{\textbf{Top:} Redshift and color bias dependence, with dashed black line at color bias of zero, where the subscript RM stands for the redMaGiC photometric redshift case. \textbf{Bottom:}  $\Lambda \textrm{CDM}$ $\mu$ bias plotted against Tripp estimator term for color, with dashed black line at x=y. Both are shown for data (\textbf{left}) and simulations (\textbf{right}). The color of each point corresponds to the redshift of the SNe as given by the colorbar on the right.}
	\label{fig:mu_c}
\end{figure*}

The formalism for uncertainty in $\mu$ is typically described as given in \citet{Kessler09}. However, in using host-galaxy photo-$z$ for this study, when naively including a $\sigma_{z}^{2}$ term in Equation \ref{eq:sig_muz}, we find that the predicted uncertainty is overestimated compared to the true measured RMS for $\mu$ residuals. We show the cause of this overestimated uncertainty in Figure \ref{fig:mu_c}. When using the photo-$z$, the dominant effect on the recovered distance is due to a change in recovered color (following Equation \ref{eq:Tripp}). While $m_B$ and $x_1$ are uncorrelated with redshift bias, the top right of Figure \ref{fig:mu_c} shows that for simulations, the color parameter and redshift bias are correlated. As shown on the bottom part of Figure \ref{fig:mu_c}, we find that the change in distance modulus $\mu$ calculated from the fiducial $\Lambda \textrm{CDM}$ model is roughly equal to the difference in color $c$ multiplied by $\beta$, denoted as $\beta \times \delta c$. The left two plots in Figure \ref{fig:mu_c} are made using the DES-5YR data, while the right two are made using our simulations. In practice, we can see on the Hubble Diagram that mis-estimated redshifts correspond to a one-to-one self correction in $\mu$ along the $\Lambda \textrm{CDM}$ curve. This trend is also observed in the data, with a redshift bias corresponding to a corrective bias in $\mu$. 

We find that including the $\sigma_{z}^{2}$ term is in general incorrect, as $\mu$ and $z$ are strongly correlated via the SALT2 color term. This change has a negligible effect on previous analyses performed with spectroscopic redshifts, as the redshift uncertainty for spec-$z$ is very close to zero. When the photometric redshift is misestimated, the light-curve fit returns a shifted color to compensate, which reduces the resulting scatter in the Hubble Diagram. 

We note that for peculiar velocities ($v_{\textrm{pec}}$) the measured distance and redshift are independent, and therefore including the $\sigma_{\textrm{vpec}}$ term as has previously been done is appropriate. While this correction empirically works for the DES redshift range, we also note it does not necessarily apply to low-$z$ SNe, which deviate further from a one-to-one correction. We further note that the relation may not be strictly linear; as the magnitude of the $z_{\textrm{redmagic}}-z_{\textrm{spec}}$ bias increases past 0.05 or more, the deviation from a strict one-to-one relation does as well.

\subsection{A2. Parent Population Parameters}
\label{sec:A2}

Here we present the parent population parameters for both the redMaGiC subset SNe and the full DES-3YR spectroscopic sample as a function of mass and parametrized as described in \citet{Popovic21}. A summary of the average population parameters (described by a mean, left-sided $\sigma_-$ and right-sided $\sigma_+$) is given in Table \ref{tab:parentpops}.

\begin{table}[ht]
  \centering
\begin{tabular}{r||ccc|ccc}
& \multicolumn{3}{c}{$c$} & \multicolumn{3}{c}{$x_1$} \\
\hline
logMass & mean & $\sigma_-$ & $\sigma_+$  & mean & $\sigma_-$ & $\sigma_+$ \\
\hline \hline
10.2  & -0.084 $\pm$ 0.053  & 0.044 $\pm$ 0.038  & 0.194 $\pm$ 0.075  & -1.798 $\pm$ 0.769   & 1.69 $\pm$ 0.821  & 2.002 $\pm$ 0.618 \\
10.4   & -0.033 $\pm$ 0.03   & 0.041 $\pm$ 0.02   & 0.03 $\pm$ 0.024  & -0.254 $\pm$ 0.615   & 1.72 $\pm$ 0.652  & 0.627 $\pm$ 0.486 \\
10.6  & -0.078 $\pm$ 0.026  & 0.024 $\pm$ 0.018   & 0.157 $\pm$ 0.04  & -0.528 $\pm$ 0.712  & 1.467 $\pm$ 0.619  & 0.808 $\pm$ 0.543 \\
10.8   & -0.07 $\pm$ 0.018  & 0.016 $\pm$ 0.012  & 0.134 $\pm$ 0.026   & -0.464 $\pm$ 0.68   & 1.512 $\pm$ 0.57  & 0.762 $\pm$ 0.534 \\
11.0  & -0.067 $\pm$ 0.018  & 0.015 $\pm$ 0.012   & 0.13 $\pm$ 0.025  & -0.453 $\pm$ 0.685   & 1.462 $\pm$ 0.55  & 0.747 $\pm$ 0.543 \\
11.2  & -0.064 $\pm$ 0.022  & 0.019 $\pm$ 0.015  & 0.134 $\pm$ 0.025   & -0.66 $\pm$ 0.741  & 2.093 $\pm$ 0.617  & 1.272 $\pm$ 0.556 \\
11.4  & -0.051 $\pm$ 0.022  & 0.018 $\pm$ 0.014  & 0.105 $\pm$ 0.023  & -1.338 $\pm$ 0.851   & 1.81 $\pm$ 0.769   & 1.742 $\pm$ 0.66 \\
11.6   & -0.021 $\pm$ 0.04  & 0.029 $\pm$ 0.023  & 0.105 $\pm$ 0.037   & -1.34 $\pm$ 0.766  & 1.574 $\pm$ 0.826     & 2.11 $\pm$ 0.6 \\
11.8    & 0.09 $\pm$ 0.098  & 0.139 $\pm$ 0.071   & 0.19 $\pm$ 0.076  & -0.875 $\pm$ 0.863  & 1.847 $\pm$ 0.776  & 2.117 $\pm$ 0.672 \\
\end{tabular}
\caption{Parent Populations for redMaGiC SN Sample}
\end{table}

\begin{table}[ht]
  \centering
\begin{tabular}{r||ccc|ccc}
& \multicolumn{3}{c}{$c$} & \multicolumn{3}{c}{$x_1$} \\
\hline
logMass & mean & $\sigma_-$ & $\sigma_+$  & mean & $\sigma_-$ & $\sigma_+$ \\
\hline \hline
8.8  & -0.087 $\pm$ 0.033  & 0.028 $\pm$ 0.025  & 0.145 $\pm$ 0.043     & 0.3 $\pm$ 0.348  & 0.374 $\pm$ 0.277   & 0.56 $\pm$ 0.234 \\
         9.0  & -0.091 $\pm$ 0.025  & 0.021 $\pm$ 0.018  & 0.139 $\pm$ 0.035    & 0.28 $\pm$ 0.312   & 0.331 $\pm$ 0.23  & 0.535 $\pm$ 0.216 \\
         9.2  & -0.094 $\pm$ 0.018  & 0.015 $\pm$ 0.013  & 0.137 $\pm$ 0.033   & 0.368 $\pm$ 0.336  & 0.451 $\pm$ 0.271   & 0.472 $\pm$ 0.22 \\
         9.4     & -0.1 $\pm$ 0.02  & 0.017 $\pm$ 0.014  & 0.169 $\pm$ 0.035   & 0.233 $\pm$ 0.338  & 0.387 $\pm$ 0.255  & 0.677 $\pm$ 0.234 \\
         9.6  & -0.097 $\pm$ 0.025  & 0.022 $\pm$ 0.017  & 0.166 $\pm$ 0.037    & 0.767 $\pm$ 0.35  & 0.925 $\pm$ 0.296  & 0.354 $\pm$ 0.243 \\
         9.8  & -0.089 $\pm$ 0.022   & 0.02 $\pm$ 0.015  & 0.158 $\pm$ 0.033   & 0.712 $\pm$ 0.369  & 1.146 $\pm$ 0.321  & 0.378 $\pm$ 0.258 \\
        10.0   & -0.09 $\pm$ 0.021  & 0.019 $\pm$ 0.014  & 0.165 $\pm$ 0.037   & 0.678 $\pm$ 0.405  & 1.332 $\pm$ 0.359  & 0.418 $\pm$ 0.274 \\
        10.2  & -0.099 $\pm$ 0.021  & 0.019 $\pm$ 0.014  & 0.204 $\pm$ 0.037    & 0.502 $\pm$ 0.51  & 1.323 $\pm$ 0.442  & 0.536 $\pm$ 0.331 \\
        10.4  & -0.091 $\pm$ 0.024  & 0.021 $\pm$ 0.016  & 0.179 $\pm$ 0.035    & 0.02 $\pm$ 0.616  & 1.204 $\pm$ 0.561  & 0.745 $\pm$ 0.407 \\
        10.6  & -0.075 $\pm$ 0.028  & 0.026 $\pm$ 0.019  & 0.147 $\pm$ 0.033  & -0.471 $\pm$ 0.546  & 0.809 $\pm$ 0.477  & 0.759 $\pm$ 0.392 \\
        10.8  & -0.087 $\pm$ 0.016   & 0.012 $\pm$ 0.01   & 0.128 $\pm$ 0.04  & -0.418 $\pm$ 0.596  & 1.456 $\pm$ 0.557  & 0.855 $\pm$ 0.453 \\
        11.0  & -0.077 $\pm$ 0.021  & 0.018 $\pm$ 0.014  & 0.133 $\pm$ 0.041  & -0.561 $\pm$ 0.692  & 1.713 $\pm$ 0.659   & 1.16 $\pm$ 0.548 \\
        11.2  & -0.062 $\pm$ 0.034  & 0.031 $\pm$ 0.025  & 0.153 $\pm$ 0.049  & -0.526 $\pm$ 0.771  & 1.702 $\pm$ 0.698  & 1.116 $\pm$ 0.604 \\
\end{tabular}
\caption{Parent Populations for DES-3YR Spectroscopic Sample}
\end{table}
\bibliographystyle{apj}
\bibliography{research2}{}

\end{document}